\documentclass[traditabstract]{aa}
\usepackage{natbib}
\usepackage{graphicx}
\usepackage{amsmath}
\usepackage{amssymb}
\usepackage[bookmarks=false, colorlinks=true, citecolor=blue, linkcolor=blue, urlcolor=blue]{hyperref}

\def\micron{\hbox{\,$\mu$m}}
\newcommand{\Herschel}{\textit{Herschel}}
\newcommand{\Spitzer}{\textit{Spitzer}}
\newcommand{\Lsun}{\hbox{$L_{\rm \odot}$}}
\newcommand{\kms}{\hbox{\hbox{km}\,\hbox{s}$^{-1}$}}

\newcommand{\FTS}{{SPIRE\slash FTS}}
\newcommand{\Hmol}{H$_2$}
\newcommand{\nht}{$n_{\rm H_2}$}
\newcommand{\tkin}{$T_{\rm kin}$}

\newcommand\nodata{ ~$\cdots$~ }

\def\CO#1{\newcount\coj
\coj=#1
\advance\coj -1
\hbox{CO\,$J=#1-\the\coj$}}

\titlerunning{Warm molecular gas in local IR bright Seyfert galaxies}
\authorrunning{Pereira-Santaella et al.}

\begin{document}

\title{Warm molecular gas temperature distribution in six local infrared bright Seyfert galaxies\thanks{\Herschel\ is an ESA space observatory with science instruments provided by European-led Principal Investigator consortia and with important participation from NASA.}}

\author{Miguel Pereira-Santaella$^{1}$, Luigi Spinoglio$^{1}$, Paul P. van der Werf$^{2}$, Javier Piqueras L\'opez$^{3}$}

\institute{Istituto di Astrofisica e Planetologia Spaziali, INAF, Via Fosso del Cavaliere 100, I-00133 Roma, Italy\\ \email{miguel.pereira@iaps.inaf.it}\label{inst1} \and 
Leiden Observatory, Leiden University, PO Box 9513, 2300, RA Leiden, The Netherlands\label{inst2} \and
Centro de Astrobiolog\'ia (INTA-CSIC), Ctra de Torrej\'on a Ajalvir, km 4, 28850, Torrej\'on de Ardoz, Madrid, Spain\label{inst3}}

\abstract{
We simultaneously analyze the spectral line energy distributions (SLEDs) of CO and \Hmol\ of six local luminous infrared (IR) Seyfert galaxies. For the CO SLEDs, we used new \Herschel\slash SPIRE FTS data (from $J=4-3$ to $J=13-12$) and ground-based observations for the lower-$J$ CO transitions. The \Hmol\ SLEDs were constructed using archival mid-IR \Spitzer\slash IRS and near-IR VLT\slash SINFONI data for the rotational and ro-vibrational \Hmol\ transitions, respectively. In total, the SLEDs contain 26 transitions with upper level energies between 5 and 15\,000\,K.
A single, constant density, model ($n_{\rm H_2}\sim 10^{4.5-6}$\,cm$^{-3}$) with a broken power-law temperature distribution reproduces well both the CO and \Hmol\ SLEDs. The power-law indices are $\beta_1\sim 1-3$ for warm molecular gas ($20$\,K$<T<100$\,K) and $\beta_2\sim 4-5$ for hot molecular gas ($T>100$\,K). We show that the steeper temperature distribution (higher $\beta$) for hot molecular gas can be explained by shocks and photodissociation region (PDR) models, however, the exact $\beta$ values are not reproduced by PDR or shock models alone and a combination of both is needed. We find that the three major mergers among our targets have shallower temperature distributions for warm molecular gas than the other three spiral galaxies. This can be explained by a higher relative contribution of shock excitation, with respect to PDR excitation, for the warm molecular gas in these mergers. For only one of the mergers, IRASF~05189--2524, the shallower \Hmol\ temperature distribution differs from that of the spiral galaxies. 
The presence of a bright active galactic nucleus in this source might explain the warmer molecular gas observed.
}


\maketitle

\begin{table*}[htp]
\caption{Sample of IR bright Seyfert galaxies observed with \Herschel\slash SPIRE FTS}
\label{tab:sample}
\centering
\begin{footnotesize}
\begin{tabular}{lccccccccccc}
\hline \hline
Galaxy Name &  R.A. & Decl. & $cz$\tablefootmark{a}  &  $D_{\rm L}$\tablefootmark{b} & Nuclear & $\log L_{\rm IR}$\tablefootmark{d} & $\alpha_{\rm AGN}$\tablefootmark{e} & Obs. ID\tablefootmark{f} \\
 &  (J2000.0) & (J2000.0) & (\kms) & (Mpc) & spect. class.\tablefootmark{c} & (\Lsun) & (\%)&\\
\hline
NGC~34 		& 00 11 06.5 & $-$12 06 26 & 5881  & 77.0 & Sy2 & 11.4 & 2 & 1342199253 \\
IRASF~05189--2524& 05 21 01.4 & $-$25 21 45 & 12760 & 181  & Sy2 & 12.2 & 71 & 1342192832, 1342192833\\
UGC~05101 	& 09 35 51.6 &   +61 21 11 & 11802 & 168  & Sy2 & 12.0 & 35 & 1342209278\\
NGC~5135 	& 13 25 44.0 & $-$29 50 01 & 4105  & 60.9 & Sy2 & 11.3 & 20 & 1342212344\\
NGC~7130 	& 21 48 19.5 & $-$34 57 04 & 4842  & 63.6 & Sy2 & 11.3 & 25 & 1342219565\\
NGC~7469 	& 23 03 15.6 &   +08 52 26 & 4892  & 56.6 & Sy1 & 11.5 & 20 & 1342199252\\
\hline
\end{tabular}
\end{footnotesize}
\tablefoot{\tablefoottext{a}{Heliocentric velocity from NASA/IPAC extragalactic database (NED).}
\tablefoottext{b}{Luminosity distance from NED.}
\tablefoottext{c}{Classification of the nuclear activity from optical spectroscopy from \citet{Yuan2010}, except for NGC~7469 \citep{AAH09PMAS}. }
\tablefoottext{d}{Logarithm of the IR luminosity, $L$(8--1000\micron), from \citet{SandersRBGS} scaled to the adopted distance.}
\tablefoottext{e}{AGN fractional contribution to the total luminosity from \citet{Esquej2012}, \citet{Veilleux2009}, and \citet{Pereira2011}.}
\tablefoottext{f}{\Herschel\ observation ID of the \FTS\ data.}
}
\end{table*}

\section{Introduction}

Molecular gas is an important phase of the interstellar medium (ISM). This phase contains a significant fraction of the total mass, and stars form in it. But the study of molecular gas presents some complications. First, the lower energy levels of \Hmol, the main component of the ISM phase, have energies $>500$\,K, thus in cold molecular gas ($T< 100$\,K) most of the \Hmol\ is in the fundamental state and no \Hmol\ emission lines are produced. And second, only the near infrared (IR) ro-vibrational \Hmol\ transitions, with $E_{\rm up}>6000$\,K, are observable from ground telescopes, so only very high-temperature ($T > 1500$\,K) molecular gas can be detected.

To overcome the first caveat, other abundant molecules with observable transitions in the millimeter range (like CO, HCN, etc.) are used as tracers of molecular gas. In particular, the lowest rotational transitions of CO, the second most abundant molecule, are commonly used to study the molecular gas content of galaxies. However, these low-$J$ CO transitions mainly originate in the coldest molecular gas. Thus, ground observations are limited to the study of either the warmest or the coldest molecular gas.

Just recently, thanks to IR and sub-millimeter space observatories like the \textit{Infrared Space Observatory} (\textit{ISO}; \citealt{Kessler1996}), the \textit{Spitzer Space Telescope} \citep{Werner2004}, and the \Herschel\ Space Observatory \citep{Pilbratt2010Herschel}, the rotational \Hmol\ transitions as well as the intermediate-$J$ CO transitions became accessible for a large number of local galaxies (e.g., \citealt{Rigopoulou02,Roussel07,vanderWerf2010,Pereira2013}). Therefore, now for the first time, it is possible to obtain a complete snapshot of molecular gas emission and study its physical properties (temperature, density, column density, etc.) and the excitation mechanisms (ultraviolet (UV) radiation, shocks, and X-ray and cosmic rays).

In this work, we present new data obtained by the Fourier transform spectrometer (FTS) module of the Spectral and Photometric Imaging Receiver (SPIRE) instrument on-board \Herschel\ \citep{Griffin2010SPIRE,Naylor2010,Swinyard2010} for six local active luminous IR galaxies. These \FTS\ data cover the 210--670\micron\ (450--1440\,GHz) spectral range, so the mid-$J$ CO lines ($J=4-3$ to $J=13-12$) are observed. We completed the CO spectral line energy distributions (SLEDs) with ground-based observations of the three lowest $J$ CO transitions.

In addition, we complemented the CO SLEDs with the \Hmol\ SLEDs obtained from near- and mid-IR observations of these galaxies. We used the available mid-IR spectroscopy obtained by the \Spitzer\ IR spectrograph (IRS; \citealt{HouckIRS}) to measure the lowest rotational \Hmol\ transitions (e.g., \citealt{Wu2009, Pereira2010}), and near-IR integral field spectroscopy obtained by the Spectrograph for INtegral Field Observations in the Near-Infrared (SINFONI; \citealt{Eisenhauer2003}) on the 
Very Large Telescope (VLT) for the ro-vibrational \Hmol\ transitions. For the first time, we have performed a radiation transfer analysis of the whole set of molecular lines together (i.e., CO rotational and \Hmol\ rotational and ro-vibrational) in local IR bright galaxies. In total, the compiled CO and \Hmol\ SLEDs contain information for 26 transitions with upper level energies between 5 and 15\,000\,K, thus the emission from most of the molecular gas is included.

The paper is organized as follows. In Section \ref{s:observations}, we present the sample and the data reduction. Sections \ref{s:rad_models} and \ref{s:sled_fit} describe the radiative transfer models used, and the fitting of the SLEDs. The cold-to-warm molecular gas ratio and the heating mechanisms are discussed in Sections \ref{s:cold-to-warm_ratio} and \ref{s:heating}, respectively. We summarize the main results in Section \ref{s:conclusions}.

\section{Observations and data reduction}\label{s:observations}

\subsection{Sample}

Our sample contains six local (50--180\,Mpc) IR bright Seyfert galaxies observed by \FTS\ through two programs (PIs: P. van der Werf and L. Spinoglio). Their $\log L_{\rm IR}\slash L_{\rm \odot}$ ranges between 11.4 and 12.2 (see Table \ref{tab:sample}). All of them host an active nucleus (AGN), although the AGN dominates the energy output of the galaxy only for IRASF~05189--2524. For the remaining galaxies, the AGN contributes between 2 and 35\,\% of the total luminosity. Three of the galaxies (NGC~34, IRASF~05189--2524, and UGC~05101) are advanced major mergers. The other three galaxies (NGC~5135, NGC~7130, and NGC~7469) are spirals, although NGC~7130 and NGC~7469 have peculiar morphologies suggesting recent minor interactions \citep{Genzel1995, Bellocchi2012}.

\subsection{\Herschel\ \FTS\ spectroscopy}\label{ss:spectroscopy}

We obtained \FTS\ high-resolution (1.45\,GHz) spectroscopic observations of six nearby Seyfert luminous IR galaxies. Integration times varied between 5 and 34 ks corresponding to a 3$\sigma$ line detection limit $\sim$0.5--1$\times10^{-15}$\,erg\,cm$^{-2}$\,s$^{-1}$.

The \FTS\ consists of two bolometer arrays, the spectrometer short wavelength (SSW; 925--1545\,GHz) and the spectrometer long wavelength (SLW; 446--948\,GHz). They sparsely cover a field of view (FoV) of $\sim$2\arcmin\ with the central bolometer centered at the nuclei of the galaxies. The full-width half-maximum (FWHM) of the SSW beam is 18\arcsec\ almost constant with frequency. For the SLW bolometers the beam FWHM varies between $\sim$30 and $\sim$42\arcsec\ with a complicated dependence on frequency \citep{Swinyard2010}.
The SSW beams size correspond to 5--15\,kpc at the distance of our galaxies, therefore these galaxies are almost point like sources for both the SSW and SLW bolometers (see also Figure 1 of \citealt{Pereira2013}).

The data was reduced as described by \citet{Pereira2013}, but using the more recent \Herschel\ interactive pipeline environment software (HIPE) version 11 \citep{Ott2010HIPE}. In brief, the pipeline first creates the interferograms from the bolometer timelines. After the interferogram phase errors are corrected and the baselines removed a Fourier transform is applied to obtain the spectra. These are dominated by the thermal emission of the telescope that is later removed. Residual background emission is subtracted by averaging the off-source bolometers. Finally, the point-source calibration is applied to the spectra extracted from the central bolometers (SLWC3 and SSWD4). The final spectra are plotted in Figure \ref{fig:spectra}.

To measure the line fluxes, we fitted a sinc function to the line profiles\footnote{The sinc function accurately models the FTS instrumental line shape.}. The local continuum was estimated using a linear fit around the line frequency. When two lines were close in frequency we fitted them simultaneously. The fluxes of the $^{12}$CO, HF, H$_2$O, and fine structure atomic transitions are listed in Table \ref{tab:lines}. Transitions of CH$^+$, H$_2$O$^+$, and OH$^+$ detected in some galaxies are reported in Table \ref{tab:lines_uncommon}.

The \FTS\ data of two of these galaxies, UGC~05101 and NGC~7130, were already presented by \citet{Pereira2013}. To take advantage of the newest pipeline and calibration they were reprocessed with the rest of the sample. For these two galaxies, the line fluxes in Tables \ref{tab:lines} and \ref{tab:lines_uncommon} agree with those in \citet{Pereira2013} within the 1$\sigma$ uncertainties.

\begin{figure*}
\centering
\includegraphics[width=1.8\columnwidth]{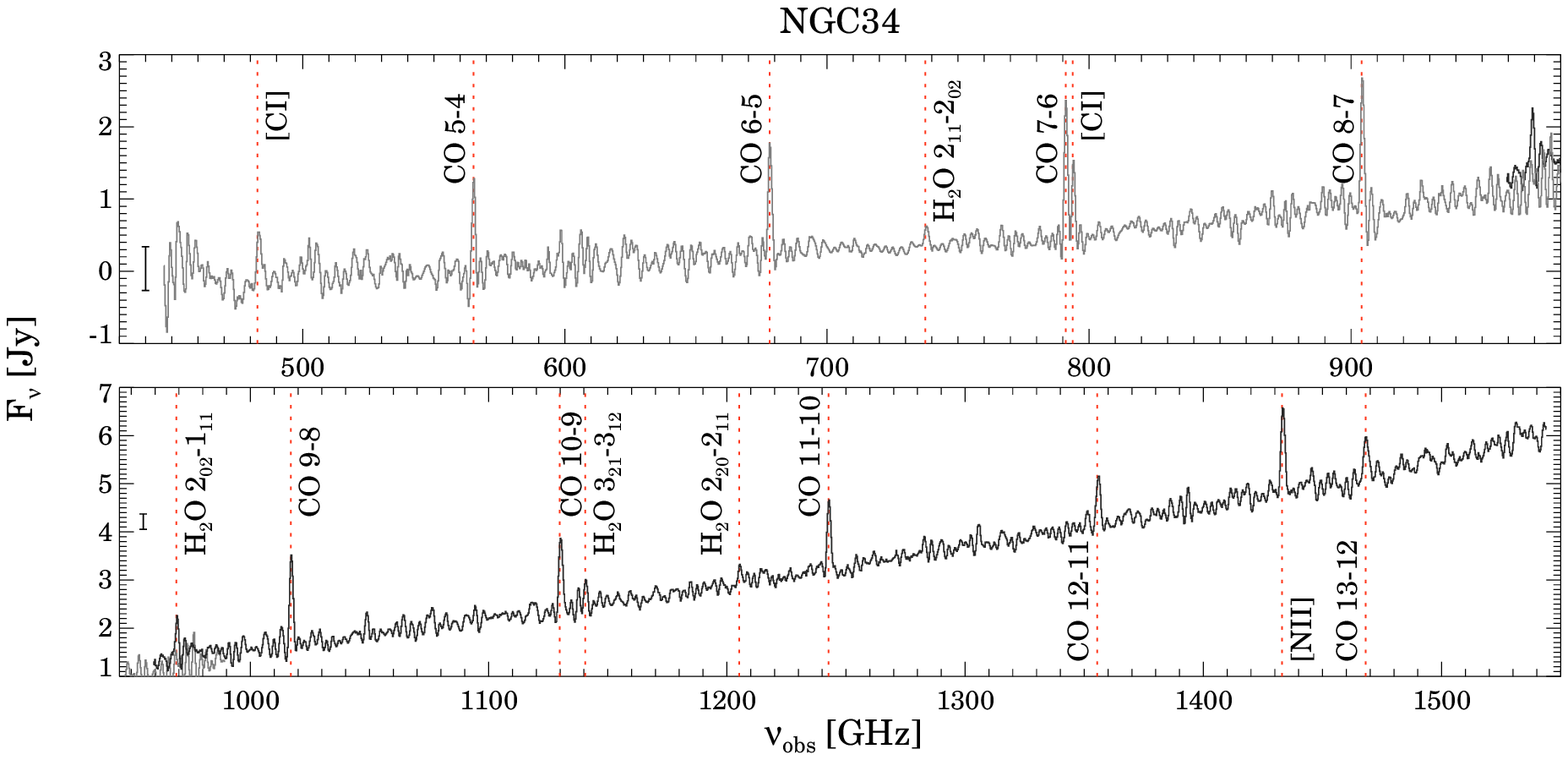}
\includegraphics[width=1.8\columnwidth]{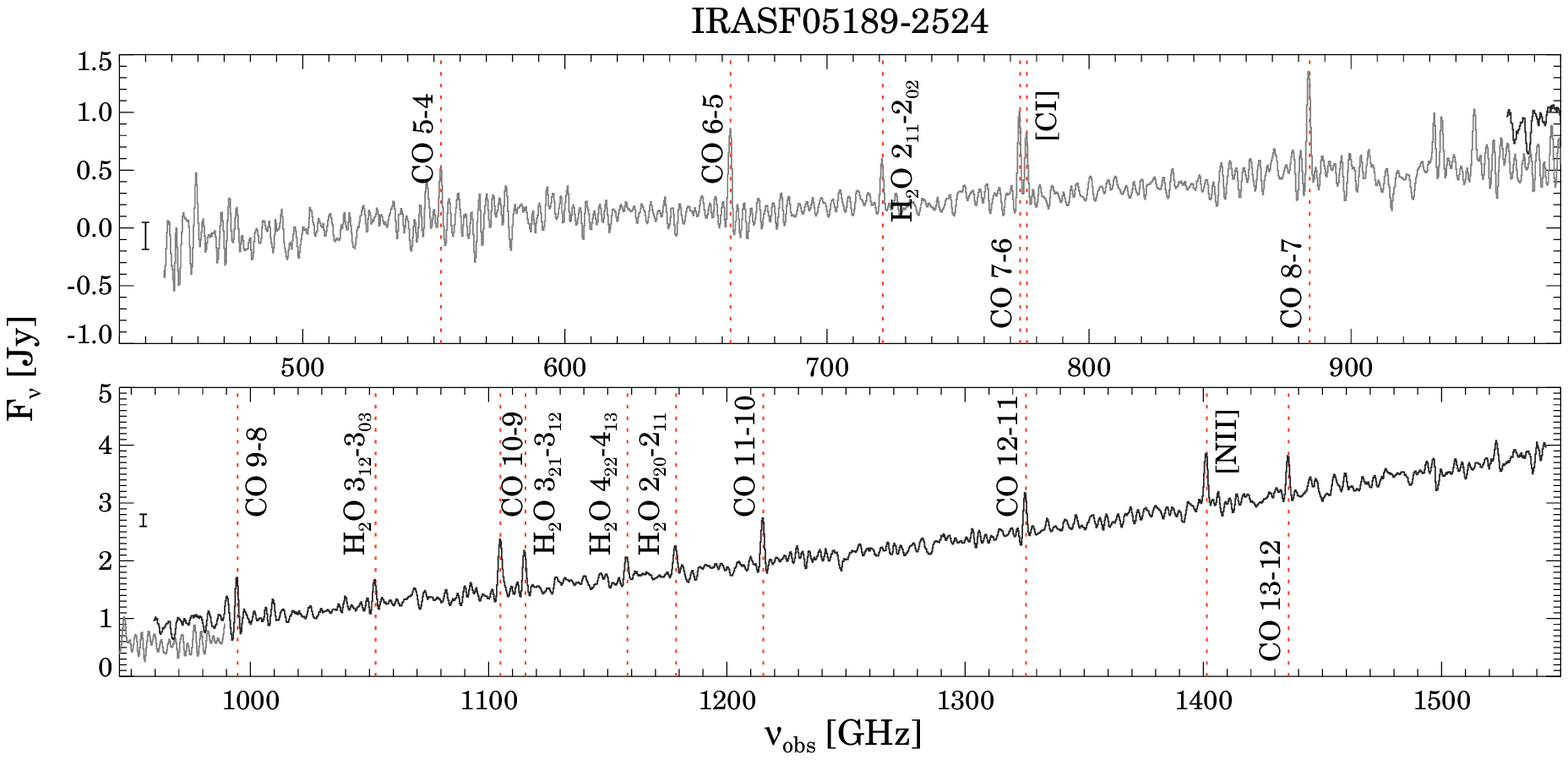}
\includegraphics[width=1.8\columnwidth]{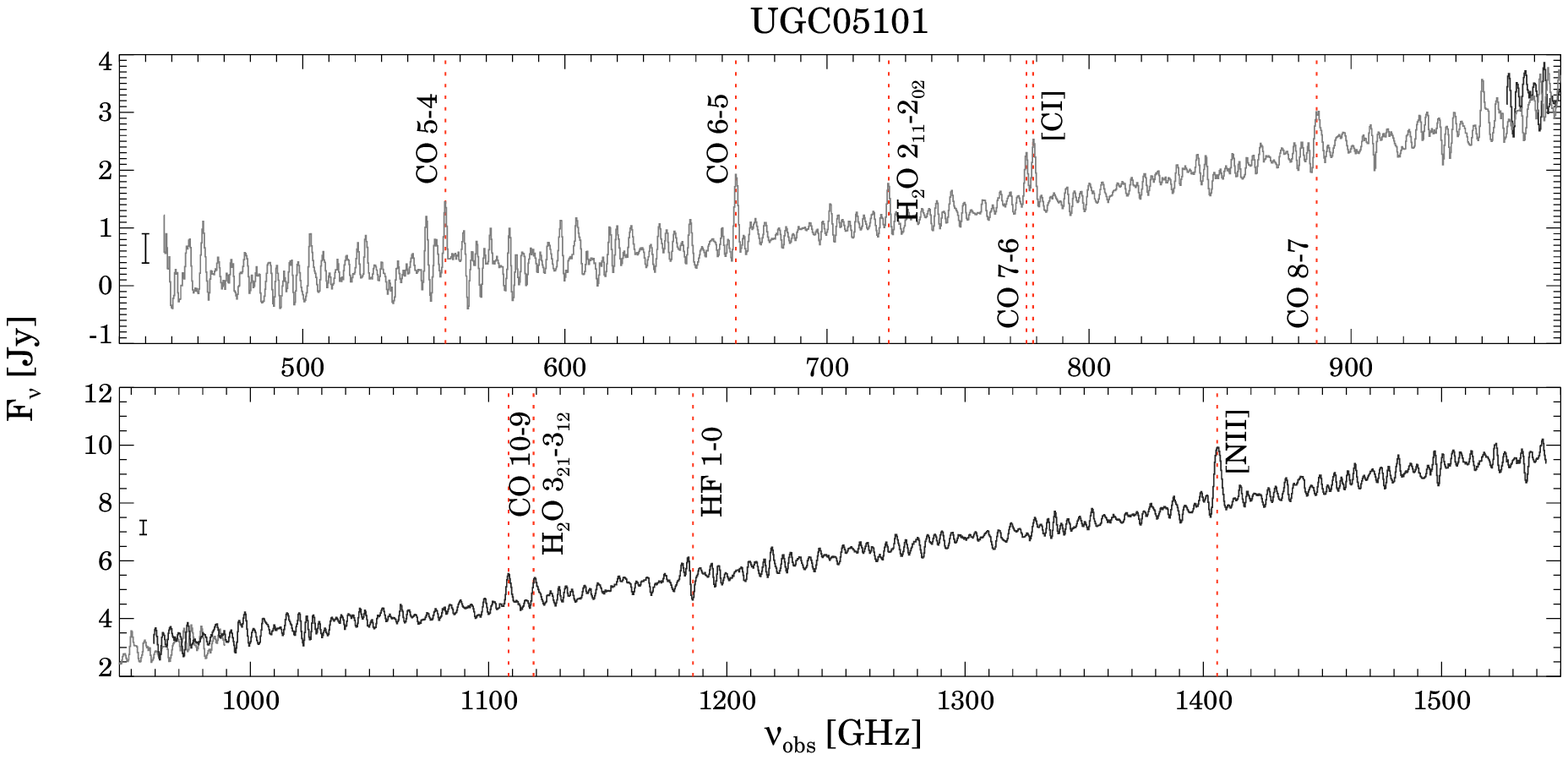}
\caption{Observed \FTS\ spectra of our sample. The black and gray lines are the SSW and SLW spectra, respectively. Notice the overlap between the two spectra in the 960 and 990\,GHz spectral range. The dashed red lines mark the position of the detected lines. The error bars indicate the median 1$\sigma$ uncertainty of each spectrum. \label{fig:spectra}}
\end{figure*}

\begin{figure*}
\addtocounter{figure}{-1}
\centering
\includegraphics[width=1.8\columnwidth]{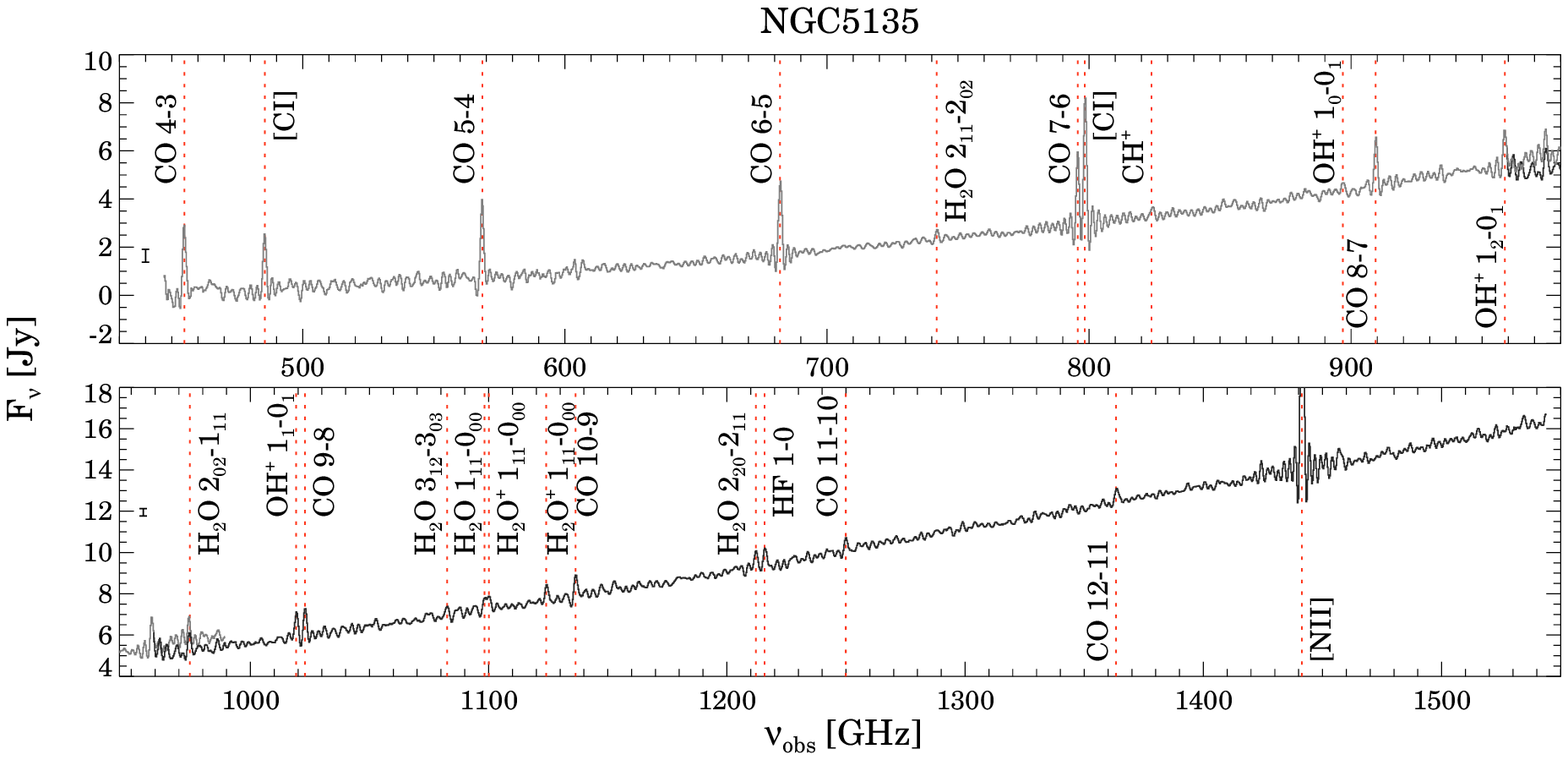}
\includegraphics[width=1.8\columnwidth]{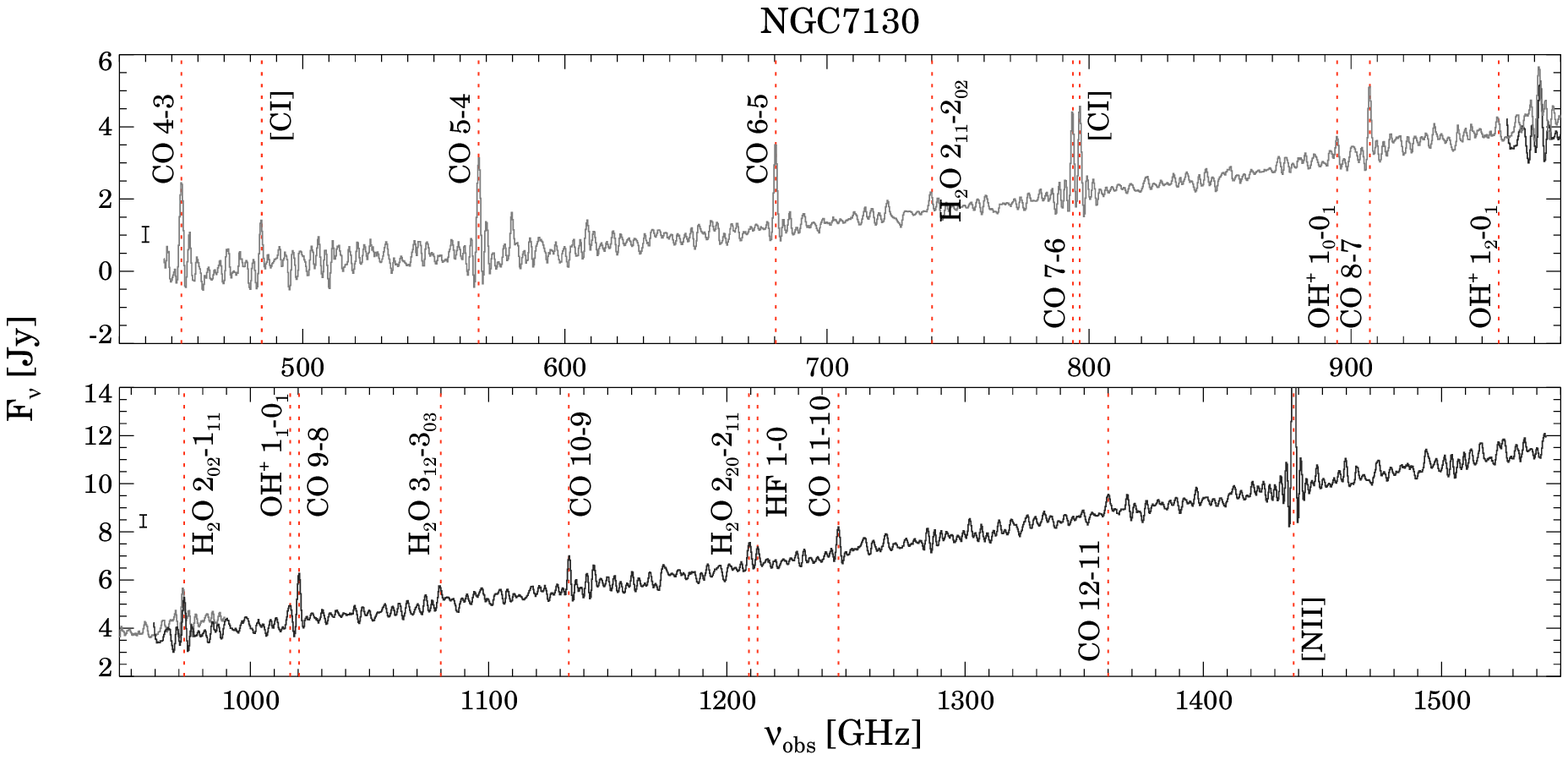}
\includegraphics[width=1.8\columnwidth]{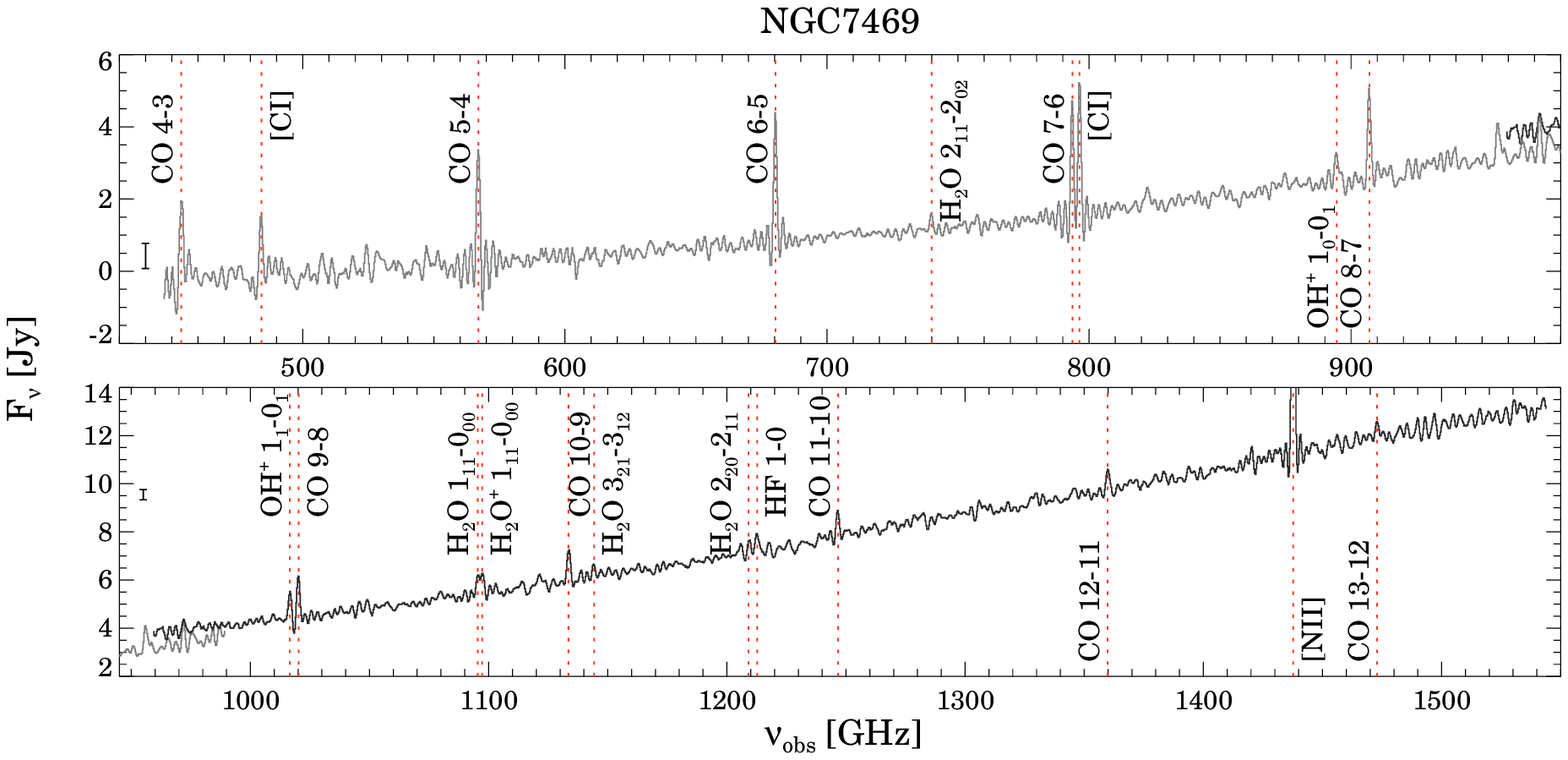}
\caption{Continued. Observed \FTS\ spectra of our sample. The black and gray lines are the SSW and SLW spectra, respectively. Notice the overlap between the two spectra in the 960 and 990\,GHz spectral range. The dashed red lines mark the position of the detected lines. The error bars indicate the median 1$\sigma$ uncertainty of each spectrum.}
\end{figure*}

\begin{table*}[ht]
\caption{SPIRE\slash FTS line fluxes}
\label{tab:lines}
\centering
\begin{small}
\begin{tabular}{lrcccccccccc}
\hline \hline
Transition &  \multicolumn{1}{c}{$\nu_{\rm rest}$}  &  & \multicolumn{6}{c}{\rm Fluxes (10$^{-15}$\,erg\,cm$^{-2}$\,s$^{-1}$)} \\
 \cline{4-9} \\[-1.5ex]
 & \multicolumn{1}{c}{(GHz)} & & NGC~34 & IRASF~05189--2524 & UGC~05101 & NGC~5135 & NGC~7130 & NGC~7469 \\
\hline
$^{12}$CO J = 4--3 & 461.041 &  &  $<$10.6 &  \nodata\tablefootmark{a}  &  \nodata\tablefootmark{a}  & 33.2 $\pm$ 3.7 & 33.0 $\pm$ 3.8 & 28.8 $\pm$ 3.3 \\
$^{12}$CO J = 5--4 & 576.268 &  & 14.3 $\pm$ 1.9 & 5.7 $\pm$ 1.3 & 11.6 $\pm$ 3.3 & 35.9 $\pm$ 3.9 & 34.2 $\pm$ 3.8 & 38.6 $\pm$ 4.2 \\
$^{12}$CO J = 6--5 & 691.473 &  & 17.7 $\pm$ 2.1 & 9.1 $\pm$ 1.3 & 14.0 $\pm$ 2.0 & 36.1 $\pm$ 3.7 & 25.6 $\pm$ 2.9 & 37.5 $\pm$ 3.9 \\
$^{12}$CO J = 7--6 & 806.652 &  & 21.3 $\pm$ 2.4 & 7.9 $\pm$ 1.1 & 10.5 $\pm$ 1.6 & 29.3 $\pm$ 3.1 & 25.5 $\pm$ 2.8 & 31.1 $\pm$ 3.3 \\
$^{12}$CO J = 8--7 & 921.800 &  & 18.2 $\pm$ 2.3 & 10.5 $\pm$ 1.6 & 10.6 $\pm$ 2.3 & 23.4 $\pm$ 2.7 & 20.8 $\pm$ 2.5 & 26.9 $\pm$ 3.1 \\
$^{12}$CO J = 9--8 & 1036.912 &  & 22.6 $\pm$ 2.6 & 8.9 $\pm$ 1.4 &  $<$10.1 & 17.5 $\pm$ 2.1 & 24.2 $\pm$ 2.8 & 21.4 $\pm$ 2.4 \\
$^{12}$CO J = 10--9 & 1151.985 &  & 18.3 $\pm$ 2.2 & 11.2 $\pm$ 1.5 & 12.1 $\pm$ 2.2 & 12.2 $\pm$ 1.8 & 16.9 $\pm$ 2.6 & 15.4 $\pm$ 2.1 \\
$^{12}$CO J = 11--10 & 1267.014 &  & 17.1 $\pm$ 2.1 & 9.3 $\pm$ 1.4 &  $<$10.0 & 7.1 $\pm$ 1.6 & 12.8 $\pm$ 2.1 & 13.0 $\pm$ 1.9 \\
$^{12}$CO J = 12--11 & 1381.995 &  & 12.5 $\pm$ 1.8 & 8.0 $\pm$ 1.2 &  $<$8.2 & 8.5 $\pm$ 1.4 & 10.5 $\pm$ 1.9 & 9.9 $\pm$ 1.7 \\
$^{12}$CO J = 13--12 & 1496.923 &  & 9.3 $\pm$ 1.7 & 8.0 $\pm$ 1.3 &  $<$10.5 &  $<$11.6 &  $<$12.1 & 9.1 $\pm$ 1.9 \\
\\
o-H$_2$O 1$_{10}$--1$_{01}$ & 556.936 &  &  $<$6.2 &  $<$5.3 &  $<$12.9 &  $<$7.6 &  $<$14.7 &  $<$12.7 \\
p-H$_2$O 2$_{11}$--2$_{02}$ & 752.033 &  & 3.6 $\pm$ 1.0 & 4.5 $\pm$ 1.0 & 7.0 $\pm$ 1.6 & 5.4 $\pm$ 1.3 & 5.9 $\pm$ 1.6 & 5.4 $\pm$ 1.4 \\
p-H$_2$O 2$_{02}$--1$_{11}$ & 987.927 &  & 9.6 $\pm$ 1.7 &  $<$6.5 &  $<$10.1 & 11.2 $\pm$ 2.5 & 18.8 $\pm$ 3.0 &  $<$8.5 \\
o-H$_2$O 3$_{12}$--3$_{03}$ & 1097.365 &  &  $<$5.3 & 5.5 $\pm$ 1.3 &  $<$9.1 & 6.3 $\pm$ 1.8 & 10.5 $\pm$ 2.3 &  $<$6.0 \\
p-H$_2$O 1$_{11}$--0$_{00}$ & 1113.343 &  &  $<$7.4 &  $<$5.7 &  $<$9.1 & 8.9 $\pm$ 1.9 &  $<$10.0 & 12.0 $\pm$ 1.6 \\
o-H$_2$O 3$_{21}$--3$_{12}$ & 1162.912 &  & 7.7 $\pm$ 1.7 & 9.1 $\pm$ 1.2 & 9.0 $\pm$ 2.1 &  $<$6.4 &  $<$12.4 & 6.8 $\pm$ 2.0 \\
p-H$_2$O 4$_{22}$-4$_{13}$ & 1207.639 &  &  $<$5.6 & 4.6 $\pm$ 1.4 &  $<$9.8 &  $<$8.0 &  $<$11.3 &  $<$7.6 \\
p-H$_2$O 2$_{20}$--2$_{11}$ & 1228.789 &  & 4.1 $\pm$ 1.1 & 4.0 $\pm$ 1.0 &  $<$9.6 & 8.9 $\pm$ 1.5 & 12.8 $\pm$ 2.1 & 6.8 $\pm$ 1.7 \\
\\ \hline \\
$[$\ion{C}{I}$]$ $^{3}$P$_{1}$-$^{3}$P$_{0}$ & 492.161 &  & 7.7 $\pm$ 2.1 &  $<$6.0 &  $<$12.6 & 28.6 $\pm$ 2.4 & 16.4 $\pm$ 3.5 & 19.9 $\pm$ 2.7 \\
$[$\ion{C}{I}$]$ $^{3}$P$_{2}$-$^{3}$P$_{1}$ & 809.342 &  & 9.9 $\pm$ 1.2 & 5.5 $\pm$ 0.7 & 12.8 $\pm$ 1.6 & 59.5 $\pm$ 1.0 & 26.7 $\pm$ 1.3 & 37.7 $\pm$ 1.3 \\ 
$[$\ion{N}{II}$]$ $^{3}$P$_{1}$-$^{3}$P$_{0}$ & 1461.132 &  & 20.7 $\pm$ 2.5 & 11.0 $\pm$ 1.4 & 31.3 $\pm$ 3.8 & 147.0 $\pm$ 3.3 & 118.3 $\pm$ 3.0 & 99.6 $\pm$ 3.7 \\
\hline
\end{tabular}
\end{small}
\tablefoot{Measured line fluxes and 1$\sigma$ statistical uncertainties. For non detections, we state the 3$\sigma$ upper limits. \tablefoottext{a} Because of the higher redshift of IRASF~05189--2524 and UGC~05101, the \CO4\ transition does not lie in the \FTS\ spectral range.}
\end{table*}
 
\begin{table*}
\caption{SPIRE\slash FTS CH$^+$, HF, o-H$_2$O$^+$, and OH$^+$ fluxes}
\label{tab:lines_uncommon}
\centering
\begin{tabular}{llrccccccccc}
\hline \hline
Galaxy & Transition & \multicolumn{1}{c}{$\nu_{\rm rest}$} & Flux \\
& & \multicolumn{1}{c}{(GHz)} &  (10$^{-15}$\,erg\,cm$^{-2}$\,s$^{-1}$) \\
\hline
UGC~05101 & HF $J = 1-0$ & 1232.476 & --10.4 $\pm$ 2.3\,$^a$ \\
\\
NGC~5135 & CH$^+$ $J = 1-0$ & 835.079 & 5.1 $\pm$ 1.6 \\
& HF $J = 1-0$ & 1232.476 & 10.5 $\pm$ 1.5 \\
 & o-H$_2$O$^+$ $1_{11}-0_{00}$ & 1115.204 & 8.8 $\pm$ 1.8      \\
 & o-H$_2$O$^+$ $1_{11}-0_{00}$ & 1139.561 & 8.9 $\pm$ 1.7 \\
 & OH$^+$ $1_{0}-0_{1}$ & 909.159  & 5.8 $\pm$ 1.6 \\
 & OH$^+$ $1_{2}-0_{1}$ & 971.805  & 23.4 $\pm$ 2.8 \\
 & OH$^+$ $1_{1}-0_{1}$ & 1032.998 & 14.9 $\pm$ 1.4 \\
\\
NGC~7130 & HF $J = 1-0$ & 1232.476 & 8.3 $\pm$ 2.0 \\
 & OH$^+$ $1_{0}-0_{1}$ & 909.159  & 6.5 $\pm$ 1.9 \\
 & OH$^+$ $1_{2}-0_{1}$ & 971.805  & 26.1 $\pm$ 3.6 \\
 & OH$^+$ $1_{1}-0_{1}$ & 1032.998 & 10.7 $\pm$ 2.2 \\
\\
NGC~7469 & HF $J = 1-0$ & 1232.476 & 7.7 $\pm$ 1.6 \\
 & o-H$_2$O$^+$ $1_{11}-0_{00}$ & 1115.204 & 12.7 $\pm$ 1.7 \\
 & o-H$_2$O$^+$ $1_{11}-0_{00}$ & 1139.561 & $<$6.8 \\
 & OH$^+$ $1_{0}-0_{1}$ & 909.159  & 8.4 $\pm$ 2.1 \\
 & OH$^+$ $1_{2}-0_{1}$ & 971.805  & $<$9.8   \\
 & OH$^+$ $1_{1}-0_{1}$ & 1032.998 & 12.7 $\pm$ 1.3 \\
\hline
\end{tabular}
\tablefoot{Measured line fluxes and 1$\sigma$ statistical uncertainties. For non detections, we list the 3$\sigma$ upper limits. $^{(a)}$ The equivalent width of the HF $J=1-0$ absorption in UGC~05101 is (3.8 $\pm$ 0.9)$\times$10$^{-6}$\,cm.}
\end{table*}

\subsection{Spitzer\slash IRS spectroscopy}\label{ss:spitzer_irs}

Mid-IR Spitzer\slash IRS spectroscopic observations of these galaxies were available on the \Spitzer\ data archive. Most of the data is already published in several papers (e.g., \citealt{Armus2004, Wu2009, Pereira2010, Tommasin2010, Esquej2012}), but the fluxes (and upper limits) of the \Hmol\ transitions are not always reported. For this reason, we retrieved the data from the archive and reprocessed it in a uniform way.

These observations include low- and high-resolution spectroscopy ($R\sim$60--120 and 600, respectively) covering the spectral range 5.5--14\micron\ (short-low resolution modules) and 10--36\micron (short- and long-high-resolution modules).

\begin{figure}
\centering
\includegraphics[width=0.95\columnwidth]{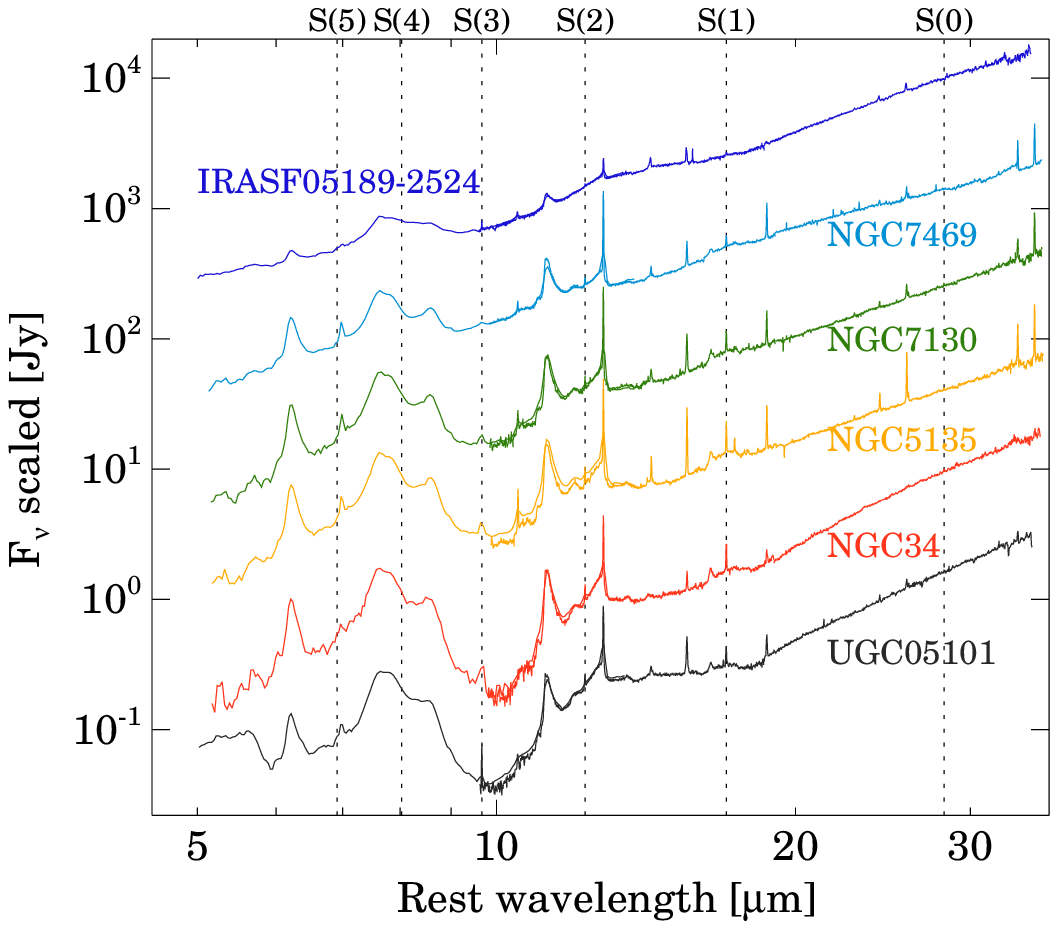}
\caption{Rest-frame \Spitzer\slash IRS spectra. The short-low (5.5--14\micron), short-high (10--19\micron), and long-high (19--36\micron) spectra are plotted for each galaxy. For clarity, the spectra are multiplied by the following factors 1, 2.5, 14, 100, 220, and 2000, from bottom to top.
The wavelengths of the \Hmol\ rotational transitions are indicated. \label{fig:spectra_irs}}
\end{figure}

The observations obtained in the staring mode were processed using the standard pipeline version S18.18 included in the Spitzer IRS Custom Extraction software (SPICE). We assumed the point source flux calibration for these galaxies. To process the spectral mapping observations, we used CUBISM \citep{SmithCUBISM}. This software combines the individual slit observations to create the spectral data cubes. From the cubes, we extracted the spectra using a 15\arcsec$\times$15\arcsec\ square aperture and then we applied an aperture correction (see \citealt{AAH2012a}). The spectra are shown in Figure \ref{fig:spectra_irs}.

We measured the fluxes of the \Hmol\ rotational transitions S(0), S(1), and S(2) in the high-resolution spectra\footnote{The \Hmol\ S(3) transition at 9.67\micron\ also lies in the high-resolution range for the higher redshift galaxies UGC~05101 and IRASF~05189-2524.} fitting a Gaussian profile and a linear function for the local continuum (Table \ref{tab:h2fluxes}).
The \Hmol\ S(3) and S(5) transitions were measured in low-resolution spectra. Because of the large number of spectral features in the mid-IR range, it is not trivial to determine the continuum for the low-resolution spectra. Therefore, we used PAHFIT \citep{Smith07} to fit the complete spectrum and estimate the continuum and line fluxes. 
It is not possible to obtain a reliable measurement of the \Hmol\ S(4) transition at 8.03\micron\ because it is blended with the 7.7\micron\ PAH broad feature.
Similarly, the \Hmol\ S(5) transition at 6.91\micron\ is blended with the [\ion{Ar}{III}]6.99\micron\ line, which in our galaxies is always stronger than the \Hmol\ line. Therefore, the \Hmol\ S(5) fluxes should be considered with caution.

The strength of the silicate feature at 9.7\micron\ ($S_{\rm Si}$) was measured in the low-resolution spectra as described by \citet{Pereira2010}. It is defined as $S_{\rm Si}=\ln \left(f_{\rm obs}\slash f_{\rm cont} \right)$, thus negative $S_{\rm Si}$ implies that the feature is seen in absorption (Table \ref{tab:extinction}). We used the strength of the silicate feature to estimate extinction in our galaxies (see Section \ref{ss:extinction}).

\subsection{SINFONI integral field spectroscopy}\label{ss:sinfoni}

We made use of K-band (1.95--2.45\,\micron) seeing-limited and adaptive optics assisted near-IR SINFONI observations for several objects of the sample. The observations where carried out during different periods, i.e., 60A (NGC~7469), 77B (NGC~5135 and NGC~7130), 80B (IRASF~05189--2524), and 82B (NGC~34), with a spectral resolution of R$\sim$4000, using different plate scales that yield different FoVs. For further details of the NGC 7469 data, see \cite{Hicks2009}, and for NGC~5135 and NGC~7130, please refer to \cite{Piqueras2012}.

We made use of archive data for IRASF~05189--2524 and NGC~34, observed on December 2007 and October 2008, respectively. The observations were made using the 0\farcs05$\times$0\farcs10\,pixel$^{-1}$ setup, which yields a FoV of $\sim$3\arcsec$\times$3\arcsec, and split into individual exposures of 900\,s following a jittering O-S-O pattern for sky and on-source frames. The total on-source integration times for each galaxy were 5400\,s for IRASF~05189--2524 and 1800\,s for NGC~34. To perform the flux calibration and to correct for the instrument response and atmospheric absorption, we used three spectrophotometric standard stars, Hip032193 and Hip052202 for IRASF~05189--2524, and Hip001115 for NGC~34, which were observed with their respective sky frames.

The reduction and calibration of the raw data were performed following the same procedure outlined in \cite{Piqueras2012}, i.e., we used the standard ESO pipeline ESOREX (version 2.0.5) to perform the standard corrections of dark subtraction, flat fielding, detector linearity, geometrical distortion, and wavelength calibration. After these corrections were applied to each frame, the sky emission was subtracted from each individual on-source exposure. We constructed individual cubes from each sky-corrected frame that were then flux-calibrated separately, and combined into a single final cube taking the relative shifts in the jittering pattern into account. The absolute flux calibration to translate the counts into physical units is based on the K-band magnitudes from the 2MASS catalog \citep{Skrutskie2006}.

\begin{figure}
\centering
\includegraphics[width=0.95\columnwidth]{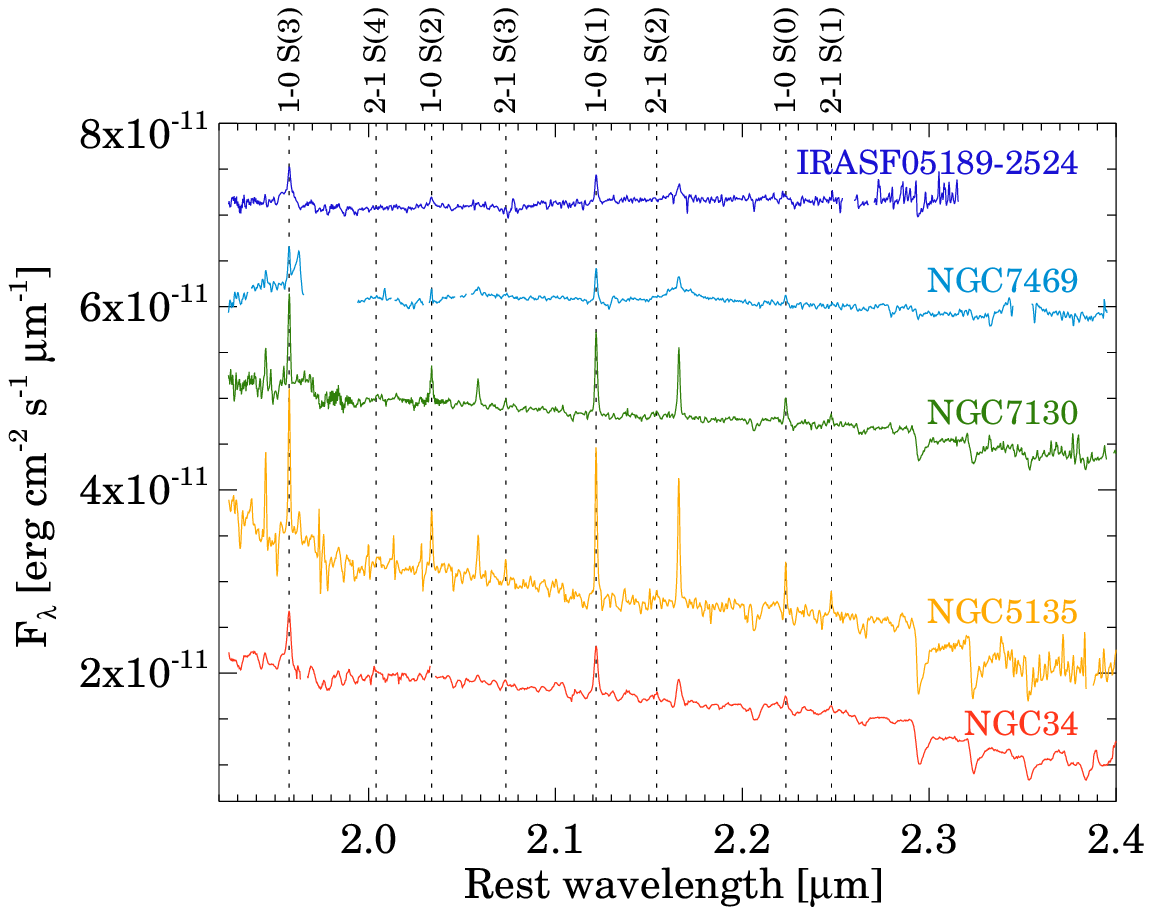}
\caption{Rest-frame SINFONI integrated spectra. For clarity the spectra are shifted by the following factors 0, 0, 3.5, 4.0, and 4.0 $\times$10$^{-11}$\,erg\,cm$^{-2}$\,s$^{-1}$ from bottom to top.
The wavelengths of the H and \Hmol\ transitions are indicated.  \label{fig:spectra_sinfoni}}
\end{figure}

To obtain the integrated K-band spectra, first we fitted the brightest \Hmol\ line (1--0 S(1) at 2.12\micron) in every spaxel to produce the intensity and velocity map of the \Hmol\ emission. Then we combined the spectra of all the spaxels with signal to noise ratio $>$ 3 in the 1--0 S(1) line, correcting for the relative velocity of each spaxel. The integrated spectra are shown in Figure \ref{fig:spectra_sinfoni}.
Finally, we measured the line fluxes and upper limits of the \Hmol\ transitions in the integrated spectra by fitting a Gaussian profile with fixed FWHM and position that were first determined from the 1--0 S(1) fit. Fluxes and upper limits are presented in Table \ref{tab:h2fluxes}.

\begin{table*}[ht]
\caption{\Hmol\ rotational and ro-vibrational fluxes}
\label{tab:h2fluxes}
\centering
\begin{small}
\begin{tabular}{lrcccccccccc}
\hline \hline
Transition &  \multicolumn{1}{c}{$\lambda$} & $A_{\rm \lambda}\slash A_{\rm K}$\tablefootmark{a} & & \multicolumn{6}{c}{\rm Fluxes (10$^{-15}$\,erg\,cm$^{-2}$\,s$^{-1}$)} \\
 \cline{5-10} \\[-1.5ex]
 & \multicolumn{1}{c}{(\micron)} & & & NGC~34 & IRASF~05189--2524 & UGC~05101 & NGC~5135 & NGC~7130 & NGC~7469 \\ 
\hline
\Hmol\ 0--0 S(0) & 28.23 & 0.42 &  & 32.2 $\pm$ 8.0 & $<$136 & 12.4 $\pm$ 4.1 & 18.0 $\pm$ 2.9 & 21.9 $\pm$ 2.9 & $<$69 \\
\Hmol\ 0--0 S(1) & 17.04 & 0.64 &  & 150 $\pm$ 12 & 25.2 $\pm$ 3.6 & 47.0 $\pm$ 7.0 & 220 $\pm$ 11 & 116.1 $\pm$ 9.8 & 182.0 $\pm$ 9.5 \\
\Hmol\ 0--0 S(2) & 12.28 & 0.58 &  & 61.7 $\pm$ 6.1 & 13.3 $\pm$ 2.5 & 22.7 $\pm$ 2.8 & 88 $\pm$ 15 & 37.7 $\pm$ 4.4 & 91.7 $\pm$ 9.2 \\
\Hmol\ 0--0 S(3) & 9.67 & 0.99 &  & 194.3 $\pm$ 8.7 & 26.3 $\pm$ 3.5 & 20.4 $\pm$ 2.3 & 233.8 $\pm$ 6.1 & 109 $\pm$ 11 & 228.5 $\pm$ 8.2 \\
\Hmol\ 0--0 S(5) & 6.91 & 0.38 &  & 87 $\pm$ 21 & $<$100 & 37.6 $\pm$ 3.4 & 66.4 $\pm$ 10.0 & 67 $\pm$ 10 & 91 $\pm$ 10 \\
\\
\Hmol\ 1--0 S(0) & 2.22 & 0.95 &  & 3.1 $\pm$ 0.2 & $<$1.3 & \nodata  & 6.4 $\pm$ 0.2 & 4.3 $\pm$ 0.1 & 1.5 $\pm$ 0.2 \\
\Hmol\ 1--0 S(1) & 2.12 & 1.02 &  & 12.4 $\pm$ 1.0 & 5.0 $\pm$ 0.4 & \nodata  & 22.2 $\pm$ 1.2 & 14.5 $\pm$ 0.6 & 5.3 $\pm$ 0.5 \\
\Hmol\ 1--0 S(2) & 2.03 & 1.09 &  & $<$7.0 & 2.0 $\pm$ 0.2 & \nodata  & 8.9 $\pm$ 0.4 & 5.6 $\pm$ 0.2 & 1.4 $\pm$ 0.2 \\
\Hmol\ 1--0 S(3) & 1.96 & 1.16 &  & 14.9 $\pm$ 0.5 & 6.9 $\pm$ 0.4 & \nodata  & 22.4 $\pm$ 0.7 & 15.8 $\pm$ 0.6 & 5.3 $\pm$ 0.2 \\
\\
\Hmol\ 2--1 S(1) & 2.25 & 0.93 &  & 1.7 $\pm$ 0.2 & $<$1.3 & \nodata  & 3.2 $\pm$ 0.2 & 1.7 $\pm$ 0.1 & $<$0.9 \\
\Hmol\ 2--1 S(2) & 2.15 & 0.99 &  & $<$1.9 & $<$0.9 & \nodata  & $<$1.6 & $<$1.2 & $<$0.8 \\
\Hmol\ 2--1 S(3) & 2.07 & 1.06 &  & $<$2.0 & $<$1.2 & \nodata  & 2.7 $\pm$ 0.4 & 1.2 $\pm$ 0.2 & $<$0.8 \\
\Hmol\ 2--1 S(4) & 2.00 & 1.11 &  & $<$2.8 & $<$1.1 & \nodata  & $<$3.2 & $<$1.7 & $<$1.5 \\
\hline
\end{tabular}
\end{small}
\tablefoot{Line fluxes and 1$\sigma$ statistical uncertainties. For non detections, we list the 3$\sigma$ upper limits. They are observed fluxes not corrected for extinction.
\tablefoottext{a}{Interpolated from the \citet{Chiar2006} local interstellar medium extinction law.}}
\end{table*}

\begin{table}[ht]
\caption{Extinction}
\label{tab:extinction}
\centering
\begin{small}
\begin{tabular}{lccccccccccc}
\hline \hline
Galaxy & $S_{\rm Si}$ & $A_{\rm K}$ \\
& & (mag) \\
\hline
NGC34 & $-$1.10 & \nodata\tablefootmark{a} \\
IRASF~05189-2524 & $-$0.36 & 0.79 \\
UGC05101 & $-$1.50 & 3.30 \\
NGC5135 & $-$0.39 & 0.86 \\
NGC7130 & $-$0.33 & 0.73 \\
NGC7469 & $-$0.13 & 0.29 \\
\hline
\end{tabular}
\end{small}
\tablefoot{$^a$ In NGC~34, the extinction toward the 10\micron\ continuum seems to be higher than that affecting the \Hmol\ emission, therefore we do not correct the fluxes of this galaxy (see Section \ref{ss:extinction}).}
\end{table}

\begin{table*}
\caption{Ground-based CO data}
\label{tab:ground_co}
\centering
\begin{small}
\begin{tabular}{lccccccccccc}
\hline \hline
Galaxy & \CO1 & Ref.\tablefootmark{a} & \CO2 & Ref.\tablefootmark{a} & \CO3 & Ref.\tablefootmark{a} \\ 
& 115.271\,GHz & & 230.538\,GHz & & 345.796\,GHz \\
\hline
NGC34 & 5.6 $\pm$ 1.1 & 1,S & 10.1 $\pm$ 2.0 & 1,S & \nodata & \nodata \\
IRASF~05189-2524 & 1.8 $\pm$ 0.5 & 3,A & 10.0 $\pm$ 2.6 & 3,I & 29.4 $\pm$ 7.2 & 3,J \\
UGC05101 & 1.8 $\pm$ 0.4 & 4,I & 9.7 $\pm$ 2.0 & 4,I & 22.5 $\pm$ 6.4 & 5,J \\
NGC5135 & 14.6 $\pm$ 3.5 & 3,I & 95 $\pm$ 21 & 3,I & 225 $\pm$ 56 & 3,J \\
NGC7130 & 12.4 $\pm$ 2.5 & 1,S & 44.8 $\pm$ 9.0 & 1,S & \nodata & \nodata \\
NGC7469 & 10.8 $\pm$ 2.6 & 2,I & 70.0 $\pm$ 16.4 & 2,I & 273.8 $\pm$ 65.2 & 2,J \\
\hline
\end{tabular}
\end{small}
\tablefoot{Fluxes and 1$\sigma$ uncertainties of ground-based observations of the low-$J$ CO lines in units of 10$^{-16}$\,erg\,cm$^{-2}$\,s$^{-1}$.
\tablefoottext{a}{Reference for the CO line flux and telescope used. (A) ARO 12\,m; (I) IRAM 30\,m; (J) JCMT 15\,m; (S) SEST 15\,m.}}
\tablebib{(1) \citet{Albrecht2007}. (2) \citet{Israel2009}. (3) \citet{Papadopoulos2012b}. (4) \citet{Pereira2013}. (5) \citet{Wilson2008}.}
\end{table*}

\subsection{Aperture matching}\label{ss:apertures}
We combined data from instruments with different FoVs and angular resolutions. The \FTS\ beam size varies between 20 and 40\arcsec, which correspond to 5--35\,kpc at the distance of our galaxies, so they appear as point-like sources at this resolution. By contrast, the \Spitzer\slash IRS slit width varies between 3.6 and 11.1\arcsec\ (1--10\,kpc), depending on the IRS module. For the most distant galaxies in our sample, IRASF~05189-2524 and UGC~05101, these apertures encompass their mid-IR emission. Likewise, most of the mid-IR emission of NGC~34 and NGC~7469 is produced in the compact nuclear region (few kpc, \citealt{Esquej2012, Diaz-Santos2010b}), thus no aperture matching corrections are needed. For the more nearby NGC~5135 and NGC~7130, \Spitzer\slash IRS spectral mapping observations are available and it is possible to extract their mid-IR spectra from a 15$\times$15\arcsec\ aperture (5$\times$5\,kpc) comparable to the \FTS\ beam size and large enough to include most of their bright nuclear mid-IR emission.

The SINFONI FoV for the IRASF~05189-2524 observations is 3$\times$3\arcsec\ (2.5$\times$2.5\,kpc). This source is unresolved in the SINFONI data (FWHM$\sim$0.3\arcsec), and therefore we expect that most of the flux is included in our measurements. For NGC~7469, the FoV of SINFONI data is 0.8$\times$0.8\arcsec\ (0.2$\times$0.2\,kpc), so only the nuclear emission is included. NGC~7469 has a kpc scale star-forming ring, thus to account for the missing near-IR H$_2$ emission we added the emission measured in the star-forming ring by \citet{DiazSantos2007} to the nuclear fluxes (Table \ref{tab:h2fluxes}). This correction increases the fluxes by a factor of $\sim$2.

Finally, the SINFONI FoVs are 8$\times$8\arcsec\, and 8$\times$16\arcsec, respectively, for NGC~5135 and NGC~7130. This corresponds to 2.5--5\,kpc which is similar to the source sizes \citep{Arribas2012}, so no correction is applied to the measured integrated SINFONI fluxes.

\subsection{Extinction correction}\label{ss:extinction}

Near- and mid-IR emission in luminous and ultra-luminous IR galaxies (LIRGs and ULIRGs) is affected by extinction due to the large amount of dust present in these galaxies. To estimate the extinction, we measured the $S_{\rm Si}$ (see Section \ref{ss:spitzer_irs}), $S_{\rm Si}$ and $A_{\rm K}$ are related according to the extinction law. In particular, using the \citet{Chiar2006} local ISM extinction law we obtained that $A_{\rm K}\slash S_{\rm Si}=-2.20$. The $S_{\rm Si}$ values in our galaxies range from $-1.50$ to $-0.13$ (Table \ref{tab:extinction}), which correspond to $A_{\rm K}$ between 0.3 and 3.3\,mag.

Regions producing \Hmol\ and $\sim$10\micron\ continuum emissions (where the $S_{\rm Si}$ is measured) may have different extinction levels. We cannot use the near-IR lines to test if we can use the estimated $A_{\rm K}$ values to correct for the obscuration effects in the \Hmol\ transitions because the differential extinction between the different transitions is small ($A_{\rm \lambda}\slash A_{\rm K}=$0.93--1.16, see Table \ref{tab:h2fluxes}). On the contrary, the \Hmol\ 0--0 S(3) transition at 9.67\micron\ lies close to the peak of the silicate absorption, so its flux is more affected by extinction than the rest of the mid-IR \Hmol\ lines. This is clear for UGC~05101, whose 0--0 S(1)\slash 0--0 S(3) ratios are 2.3$\pm$0.4 and 0.8$\pm$0.1 before and after the extinction correction. Actually, all the galaxies (except NGC~34, see below) have very similar 0--0 S(1)\slash 0--0 S(3) ratios, 0.76$\pm$0.06, after the extinction correction based on $S_{\rm Si}$.

NGC~34 has a strong silicate absorption ($S_{\rm Si}=-1.10$) and an observed 0--0 S(1)\slash 0--0 S(3) $=0.77\pm0.07$, which is compatible with the average extinction corrected ratio of the other galaxies. Therefore, it is possible that the regions producing the \Hmol\ and $\sim$10\micron\ continuum emissions in NGC~34 suffer from very different extinction levels. Moreover, the NGC~34 uncorrected \Hmol\ SLED does not have any indication of attenuation of the 0--0 S(3) transition (see Figure \ref{fig:powerlaw_fit}). This is at odds with the UGC~05101 ($S_{\rm Si}$=--1.50) SLED that shows a clear deficit of 0--0 S(3) emission. As a consequence, we do not apply any correction to NGC~34 since we do not have an accurate measurement of extinction toward the \Hmol\ emitting regions.

\section{Radiative transfer models}\label{s:rad_models}

Previous works have analyzed the \Herschel\ mid-$J$ CO SLEDs of galaxies by fitting non-LTE radiative transfer models  (e.g., \citealt{Rangwala2011, Spinoglio2012, Pereira2013, Rigopoulou2013}). In general, a model with two components at different temperatures is used. In these models, the ``cold'' component (usually $T_{\rm kin}<100$\,K) reproduces the lowest $J$ CO lines, whereas the ``warm'' component ($T_{\rm kin}$ several hundred K) accounts for the higher $J$ emission.

Similarly, when the mid-IR rotational \Hmol\ transitions are analyzed, models with two components are used (e.g., \citealt{Rigopoulou02, Roussel07}). In these studies, the derived $T_{\rm kin}$ are always higher than $\sim$100\,K since at lower temperatures the rotational \Hmol\ emission is negligible. In addition, it is assumed that \Hmol\ is in LTE because the critical density of these lines is low ($n_{\rm H_2}< 10^4$\,cm$^{-3}$, see Figure 3 of \citealt{Roussel07}).

For the near-IR ro-vibrational \Hmol\ lines, the molecular gas is usually assumed LTE too, although the critical densities of these transitions are higher ($n_{\rm crit}>$10$^{7}$\,cm$^{-3}$). The derived temperatures from these lines are around 2000\,K. In general, it is concluded that they are collisionally excited and UV excitation has relatively low importance for the excitation of the \Hmol\ $\nu=1-0$ transitions (e.g., \citealt{Davies2003, Bedregal2009}).

In this work, we analyze the SLED of low- and mid-$J$ CO transitions as well as pure rotational and ro-vibrational \Hmol\ transitions. Therefore, to explain the emission of all these lines, our model would need at least four components with different temperatures ranging from $\sim$50\,K to 2000\,K. 
Instead, we use an alternative approach assuming a continuous distribution of temperatures. This is advantageous since we avoid using a somewhat arbitrary number of model components to fit the data.

It is common to find that the observed SLEDs have a positive curvature, that is, the rotational temperature derived from a pair of consecutive transitions increases with the energy of their upper levels (e.g., \citealt{Neufeld2012}). Therefore, emission could come from gas with a continuous distribution of temperatures instead of several discrete temperatures. To model this distribution of temperatures we assumed that the column density follows a power-law (d$N\sim T^{-\beta}$d$T$). A similar method has been used to model the ro-vibrational and rotational \Hmol\ emission in shocked gas (e.g., \citealt{Brand1988, Neufeld2008, Shinn2009}), or the CO emission (e.g., \citealt{Goicoechea2013}).

In the following, we describe our model in detail. First, we explain how the models for single temperature and density are obtained, and then how they are combined to reproduce a power-law temperature distribution.

\subsection{Single temperature non-LTE model}\label{ss:single_model}

To solve the radiative transfer equations for the molecular emission we used the code RADEX \citep{vanderTak2007}. This code uses the escape probability approximation to obtain the molecular level populations and the intensities of the emission lines.

We constructed two sets of models, one for CO and other for \Hmol, covering a wide range of kinetic temperature ($T_{\rm kin}=10-2800$\,K) and \Hmol\ density ($n_{\rm H_2} = 10^2-10^9$\,cm$^{-3}$). For the \Hmol\ grid, we also consider collisions with atomic H with $n_{\rm H}/n_{\rm H_2}$ ratios between 1 and 10$^{-5}$. In these grids, we assume the LTE \Hmol\ ortho-to-para ratio, which varies between 0 for low temperatures and 3 for $T_{\rm kin}>200$\,K (see Figure 4 of \citealt{Burton1992}).

The \Hmol\ transitions are optically thin for $N_{\rm H_2}\slash \Delta v$ lower than $\sim$10$^{24}$\,cm$^{-2}$(\kms)$^{-1}$, thus in our models we assume optically thin \Hmol\ emission\footnote{$\Delta v\sim 100$\,\kms in our galaxies.}. Similarly, we assume optically thin mid-$J$ CO emission, this condition is fulfilled when $N_{\rm CO}\slash \Delta v$ is lower than $\sim$10$^{15}$\,cm$^{-2}$(\kms)$^{-1}$. In general, this is the case for the mid-$J$ CO observations of galaxies (e.g., \citealt{Kamenetzky2012, Pereira2013}).

For the CO grid, we used the collisional rate coefficients of \citet{Yang2010} expanded by \citet{Neufeld2012} for high-$J$ levels ($J=41$ to 80). This is important for the high temperature models as a non negligible fraction of the CO molecules are in energy levels with $J>40$ that otherwise are ignored.

We created two sub-grids of \Hmol\ models. One for high-temperature ($T_{\rm kin}> 100$\,K) and another for low temperature. For the high-temperature sub-grid we used the \Hmol-\Hmol\ collisional rate coefficients of \citet{LeBourlot1999} and the \Hmol-H coefficients of \citet{Wrathmall2007} that cover kinetic temperatures between 100 and 6000\,K. For the lower temperature sub-grid ($T_{\rm kin}< 100$\,K) we used the \Hmol-\Hmol\ collisional coefficients of \citet{Lee2008}. They only include the lowest nine rotational energy levels of \Hmol\ ($E_{\rm upper}<5900$\,K), but at $T_{\rm kin}<$100\,K, no molecules are expected at higher rotational energy levels. Collisions with atomic H are mostly important for vibrational excited \Hmol\ levels, so they are neglected for the low-temperature sub-grid. The grids for ortho-\Hmol\ and para-\Hmol\ were constructed separately, although they were later combined assuming the LTE \Hmol\ ortho-to-para ratio.

\subsection{Power-law temperature distribution}\label{ss:power_law}

We want to obtain the level populations for gas following a power-law temperature distribution, d$N\sim T^{-\beta}$d$T$. From the grids described in Section \ref{ss:single_model}, we can obtain the fractional population of the molecular level $i$, $n_i\left(T, n_{\rm H_2}\right)$, for a given kinetic temperature and \Hmol\ density. Therefore, the total column density of molecules in the energy level $i$ for a kinetic temperature distribution following a power-law between $T_0$ and $T_1$ is
\begin{equation}
N_i\left(\beta, n_{\rm H_2}\right) = A \int_{T_ 0}^{T_1}{T^{-\beta}n_i\left(T, n_{\rm H_2}\right){\rm d}T},
\end{equation}
where $A=N (\beta - 1)\slash (T_0^{1 - \beta} - T_1^{1 - \beta})$ is a normalization constant and $N$ is total molecular gas column density\footnote{By definition $\sum_i{n_i\left(T, n_{\rm H_2}\right)} = 1$, and $\sum_i{N_i\left(\beta, n_{\rm H_2}\right)}=N$, then $N = A\int_{T_0}^{T_1}{T^{-\beta}{\rm d}T}$ and $A$ can be solved.}.
In our models, we use $T_0=20$\,K and $T_1=3500$\,K. This temperature range is required to model the lower-$J$ CO and the near-IR \Hmol\ transitions, respectively.

In Section \ref{s:sled_fit}, we show that it is not possible to fit simultaneously the SLEDs of both CO and \Hmol\ using a single power-law temperature distribution. We find that the temperature distribution is much steeper for \Hmol\ ($\beta\sim 5$) than for CO ($\beta\sim 3$). Therefore, for the SLED fitting we used a broken power-law model with two exponents ($\beta_1$ and $\beta_2$). This can be modeled with the following equation:

\begin{align}
N_i\left(\beta, n_{\rm H_2}\right) & = B \left( \int_{T_0}^{T_{\rm b}}{T^{-\beta_1}n_i\left(T, n_{\rm H_2}\right){\rm d}T} \right. \nonumber \\ 
 & + \left. T_{\rm b}^{\beta_2 - \beta_1}\int_{T_{\rm b}}^{T_1}{T^{-\beta_2}n_i\left(T, n_{\rm H_2}\right){\rm d}T} \right)
\end{align}

where $T_{\rm b}$ is the break temperature, and $B= \scriptstyle{N\left((T_{\rm b}^{1-\beta_1} - T_0^{1-\beta_1})\slash (1 - \beta_1) + T_{\rm b}^{\beta_2 - \beta_1}(T_1^{1-\beta_2} - T_{\rm b}^{1-\beta_2})\slash (1 - \beta_2)\right)^{-1}}$.

\section{CO and \Hmol\ SLED fitting}\label{s:sled_fit}

To fit the SLEDs, we converted the observed fluxes into beam averaged column densities using the following relation
\begin{equation}
\langle N_i \rangle = \frac{4\pi}{\Omega} \frac{F_{ij}}{h \nu_{ij} A_{ij}}
\end{equation}

where $F_{ij}$ is the flux of the transition between levels $i$ and $j$, $\nu_{ij}$ and $A_{ij}$ are its frequency and Einstein coefficient, respectively, $h$ is the Planck constant, and $\Omega$ is the beam solid angle. Since our sources are barely resolved by \Herschel\ we assume a beam FWHM of 18\arcsec, which is the SSW beam size.
This beam size is also used for the rest of measurements as they should include most of the emission in this area (see Section \ref{ss:apertures} for details).
Using this relation, we assume that the emission is optically thin, which is reasonable for the \Hmol\ transitions as well as for the mid-$J$ CO lines (see Section \ref{s:rad_models}).

First, we attempted to fit the SLEDs using a power-law temperature distribution. The free parameters of this model are the CO and \Hmol\ column densities, $n_{\rm H_2}$ and $n_{\rm H}$ densities, and $\beta$. However, it was not possible to obtain a good fit to both the CO and \Hmol\ emissions. Nevertheless, the CO and \Hmol\ SLEDs individually are well reproduced by a power-law model, although each SLED has different $\beta$ values. 

\begin{figure}
\centering
\includegraphics[width=0.85\columnwidth]{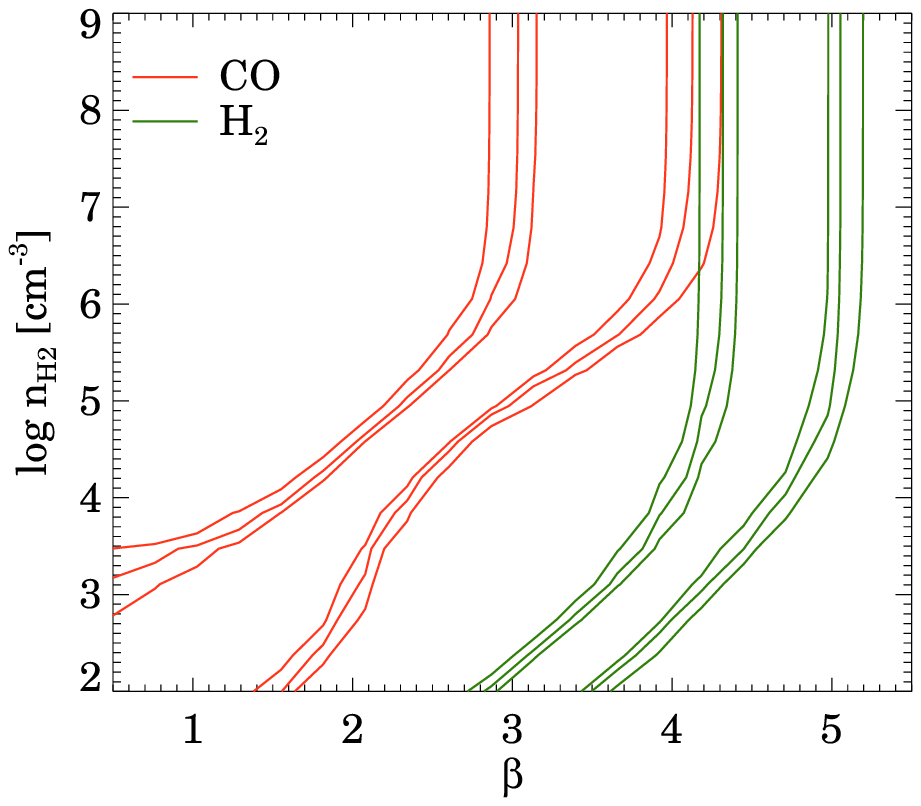}
\caption{$\chi^2$ values for the power-law temperature distribution model as a function of $\beta$ and $n_{\rm H_2}$ for NGC~7130. The red (green) contours are the 1, 2, and 3$\sigma$ confidence levels for the CO (\Hmol) SLED.\label{fig:chi_powerlaw}}
\end{figure}

In these models, there is a strong degeneracy between $\beta$ and $n_{\rm H_2}$, $\beta$ decreases for decreasing $n_{\rm H_2}$ when densities are below the critical density of the transitions ($n_{\rm H_2}<10^6$\,cm$^{-3}$ for CO and $n_{\rm H_2}<10^4$\,cm$^{-3}$ for \Hmol). This is because changes in $n_{\rm H_2}$ and $\beta$ produce similar variations in the SLED shape when the gas is not thermalized. In Figure \ref{fig:chi_powerlaw}, we plot the $\chi^2$ values for one of the galaxies where this degeneracy is evident. In addition, it is clear that the CO and \Hmol\ SLEDs are not simultaneously reproduced by any pair of $\beta$ and $n_{\rm H_2}$ parameters. For our galaxies, we find that $\beta$ is always higher for \Hmol\ (steeper temperature distribution) for any given $n_{\rm H_2}$. The $\beta$ values range between 2.5 and 3.5 for CO and between 4.5 and 5.5 for \Hmol\ for $n_{\rm H_2}$ higher than the critical densities of the transitions.

As anticipated in Section \ref{ss:power_law}, to account for the difference on the $\beta$ values for the CO and \Hmol\ SLEDs, we used an alternative model consisting of a broken power-law, which has two exponent values, $\beta_1$ and $\beta_2$, for temperatures below and above the threshold temperature $T_{\rm b}$, respectively. With this model, it is possible to fit all the available CO and \Hmol\ transitions for our galaxies, with the exception of the \CO1\ transition, which is always underpredicted by the model. The later can be caused by the presence of low temperature molecular gas that would not follow the power-law distribution. Consequently, we included a cold gas component at 15\,K assuming optically thin emission for the \CO1\ transition. In general, \CO1\ emission is not optically thin, so the obtained column densities are lower limits.

\begin{table}
\centering
\rotatebox{90}{
\begin{minipage}{\textheight}
\centering
\caption{Best-fit parameters of the CO and \Hmol\ SLEDs.}
\label{tab:mol_gas}
\begin{small}
\begin{tabular}{lccccccccccccc}
\hline \hline
Galaxy & $\beta_1$\tablefootmark{a} & $\beta_2$\tablefootmark{b} & $\log N_{\rm w}({\rm CO})$\tablefootmark{c} & $\log N_{\rm w}({\rm H_2})$\tablefootmark{d} & $\log x_{\rm CO}$\tablefootmark{e} &  $\log N_{\rm c}({\rm CO})$\tablefootmark{f} & $\log n_{\rm H}$\tablefootmark{g} & $\chi^2_{\rm red}$\tablefootmark{h} & $\log N_{\rm c}({\rm H_2})$\tablefootmark{i} & $\log N_{\rm c}({\rm H_2})\slash N_{\rm w}({\rm H_2})$ \\
 & & & (cm$^{-2}$) & (cm$^{-2}$) &  & (cm$^{-2}$) & (cm$^{-3}$) & & (10$^{21}$ cm$^{-2}$)  \\
\hline
NGC~34 		 & 0.0 $\pm$ 1.8 & 4.2 $\pm$ 0.4 & 16.0 $\pm$ 0.1 & 20.4 $\pm$ 0.2 & --4.4 $\pm$ 0.3 & 16.5 $\pm$ 0.3 & 2.9 $\pm$ 0.4 & 3.8 & 22.1 $\pm$ 0.1 & 1.7 $\pm$ 0.3 \\
 		 & 2.2 $\pm$ 0.2 & 4.9 $\pm$ 0.3 & 15.9 $\pm$ 0.1 & 21.6 $\pm$ 0.2 & --5.7 $\pm$ 0.3 & 16.5 $\pm$ 0.2 & 3.5 $\pm$ 0.3 & \bf 1.0 &                & 0.5 $\pm$ 0.3 \\
IRASF~05189-2524 & 0.0 $\pm$ 2.1 & 3.4 $\pm$ 0.5 & 15.6 $\pm$ 0.1 & 19.5 $\pm$ 0.2 & --3.9 $\pm$ 0.3 & 16.2 $\pm$ 0.3 & 3.0 $\pm$ 0.3 & 3.8 & 21.6 $\pm$ 0.2 & 2.1 $\pm$ 0.4 \\ 
		 & 2.2 $\pm$ 0.1 & 4.3 $\pm$ 0.2 & 15.6 $\pm$ 0.1 & 20.8 $\pm$ 0.1 & --5.2 $\pm$ 0.2 & 16.2 $\pm$ 0.1 & 3.5 $\pm$ 0.2 & \bf 0.4 &                & 1.0 $\pm$ 0.3 \\ 
UGC~05101	 & 0.4 $\pm$ 2.4 & 4.4 $\pm$ 0.1 & 15.8 $\pm$ 0.5 & 20.9 $\pm$ 0.7 & --5.1 $\pm$ 0.7 & 16.1 $\pm$ 0.6 & \nodata       & \bf 15.6 & 21.6 $\pm$ 0.2 & 0.7 $\pm$ 0.9 \\
 	 	 & 2.2 $\pm$ 1.5 & 4.7 $\pm$ 0.2 & 15.7 $\pm$ 0.5 & 21.9 $\pm$ 1.0 & --6.2 $\pm$ 1.0 & 16.1 $\pm$ 0.7 & \nodata       & 22.9 &                & --0.3 $\pm$ 1.0 \\
NGC~5135	 & 1.5 $\pm$ 0.3 & 4.5 $\pm$ 0.2 & 16.5 $\pm$ 0.1 & 21.4 $\pm$ 0.2 & --5.0 $\pm$ 0.3 & 17.1 $\pm$ 0.2 & 3.6 $\pm$ 0.2 & \bf 2.2 & 22.3 $\pm$ 0.1 & 0.9 $\pm$ 0.3 \\
 		 & 3.6 $\pm$ 0.3 & 4.7 $\pm$ 0.2 & 16.5 $\pm$ 0.1 & 22.9 $\pm$ 0.3 & --6.4 $\pm$ 0.4 & 16.8 $\pm$ 0.3 & 3.7 $\pm$ 0.4 & 3.3 &                & --0.6 $\pm$ 0.4 \\
NGC~7130 	 & 0.9 $\pm$ 0.5 & 4.7 $\pm$ 0.2 & 16.3 $\pm$ 0.2 & 20.9 $\pm$ 0.2 & --4.6 $\pm$ 0.3 & 17.0 $\pm$ 0.2 & 3.8 $\pm$ 0.3 & 2.2 & 22.4 $\pm$ 0.1 & 1.5 $\pm$ 0.3 \\
 		 & 3.3 $\pm$ 0.2 & 4.9 $\pm$ 0.1 & 16.4 $\pm$ 0.1 & 22.4 $\pm$ 0.1 & --6.0 $\pm$ 0.2 & 16.9 $\pm$ 0.1 & 4.0 $\pm$ 0.2 & \bf 0.9 &                & 0.0 $\pm$ 0.2 \\
NGC~7469	 & 1.2 $\pm$ 0.3 & 4.3 $\pm$ 0.2 & 16.4 $\pm$ 0.1 & 21.0 $\pm$ 0.2 & --4.6 $\pm$ 0.3 & 17.0 $\pm$ 0.2 & 3.0 $\pm$ 0.2 & 1.4 & 22.8 $\pm$ 0.1 & 1.8 $\pm$ 0.3 \\
 		 & 3.4 $\pm$ 0.1 & 4.9 $\pm$ 0.2 & 16.5 $\pm$ 0.1 & 22.6 $\pm$ 0.2 & --6.1 $\pm$ 0.3 & 16.9 $\pm$ 0.2 & 2.7 $\pm$ 0.8 & \bf 0.9 &                & 0.2 $\pm$ 0.3 \\
\hline
\end{tabular}
\end{small}
\tablefoot{The first and second line for each galaxy correspond to the best-fit parameters with fixed $n_{\rm H_2}=10^{4.5}$\,cm$^{-3}$ and $n_{\rm H_2}=10^6$\,cm$^{-3}$, respectively.
Column densities are calculated for a circular beam of 18\arcsec of diameter.
\tablefoottext{a,b}{$\beta_1$ and $\beta_2$ are the power-law indices for the CO and \Hmol\ emissions respectively.}
\tablefoottext{c,d}{Warm CO and \Hmol\ column densities respectively.}
\tablefoottext{e}{CO abundance relative to \Hmol\ in the warm molecular gas.}
\tablefoottext{f}{Cold CO column density. This value is a lower limit since we assumed optically thin emission for the \CO1\ transition.}
\tablefoottext{g}{Atomic H density derived from the ro-vibrational \Hmol\ transitions.}
\tablefoottext{h}{Reduced $\chi^2$ of the fit. The best $\chi^2$ is marked in boldface in the table.}
\tablefoottext{i}{Cold \Hmol\ column density derived from the \CO1\ intensity. We used the conversion factor $N_{\rm H_2}$(cm$^{-2}$) = $2\times 10^{20} I_{{\rm CO} J =1-0}$\,(K\,\kms) \citep{Leroy2011}.}}
\end{minipage}}
\end{table}

The value of the breaking temperature ($T_{\rm b}$) is not well constrained because the upper level energies of the CO and \Hmol\ transitions do not overlap. When $T_{\rm b}$ is left as a free parameter it varies between 100 and 500\,K with 1$\sigma$ uncertainties $\sim$100\,K, and its variation affects the $\beta$ values slightly. The average $T_{\rm b}$ is 200$\pm$130\,K. It is not possible to accurately determine $T_{\rm b}$ with our data, therefore, we fixed it to the average value.

As discussed above, $\beta_1$ and $n_{\rm H_2}$ are strongly correlated, specially for CO, when $n_{\rm H_2}$ is below the critical densities. The CO column density ($N_{\rm CO}$) is relatively constant when $\beta_1$ and $n_{\rm H_2}$ vary. But, on the contrary, the \Hmol\ column density variation is large. This is because we only measure the warm \Hmol\ column density directly, which represents a small fraction of the total \Hmol, and the total \Hmol\ column density is extrapolated using the $\beta_1$ exponent. Therefore, in our model, lower $n_{\rm H_2}$ also implies lower \Hmol\ column densities (and higher CO abundances) because of the smaller value of $\beta_1$.

\begin{figure*}
\centering
\rotatebox{90}{
\begin{minipage}{\textheight}
\centering
\includegraphics[width=0.49\textheight]{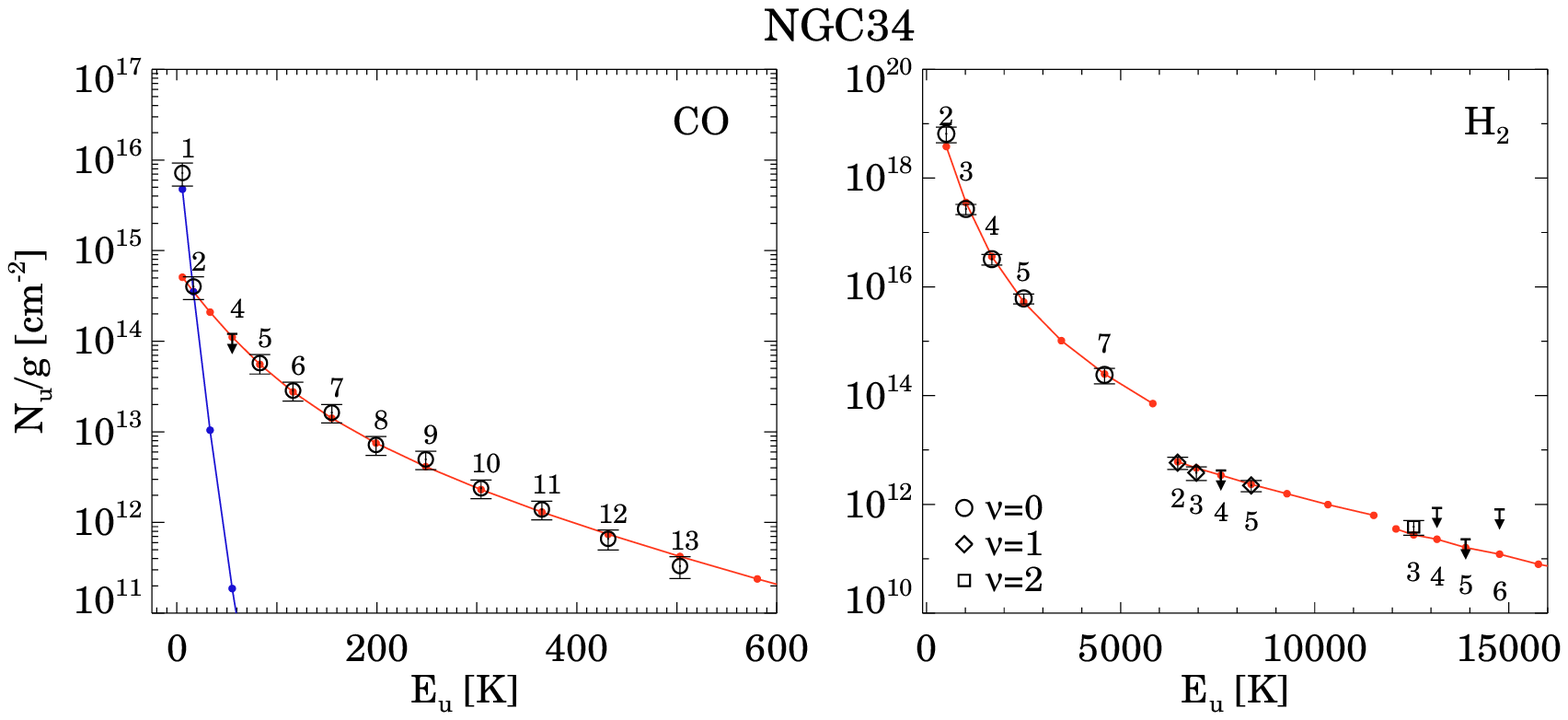}
\includegraphics[width=0.49\textheight]{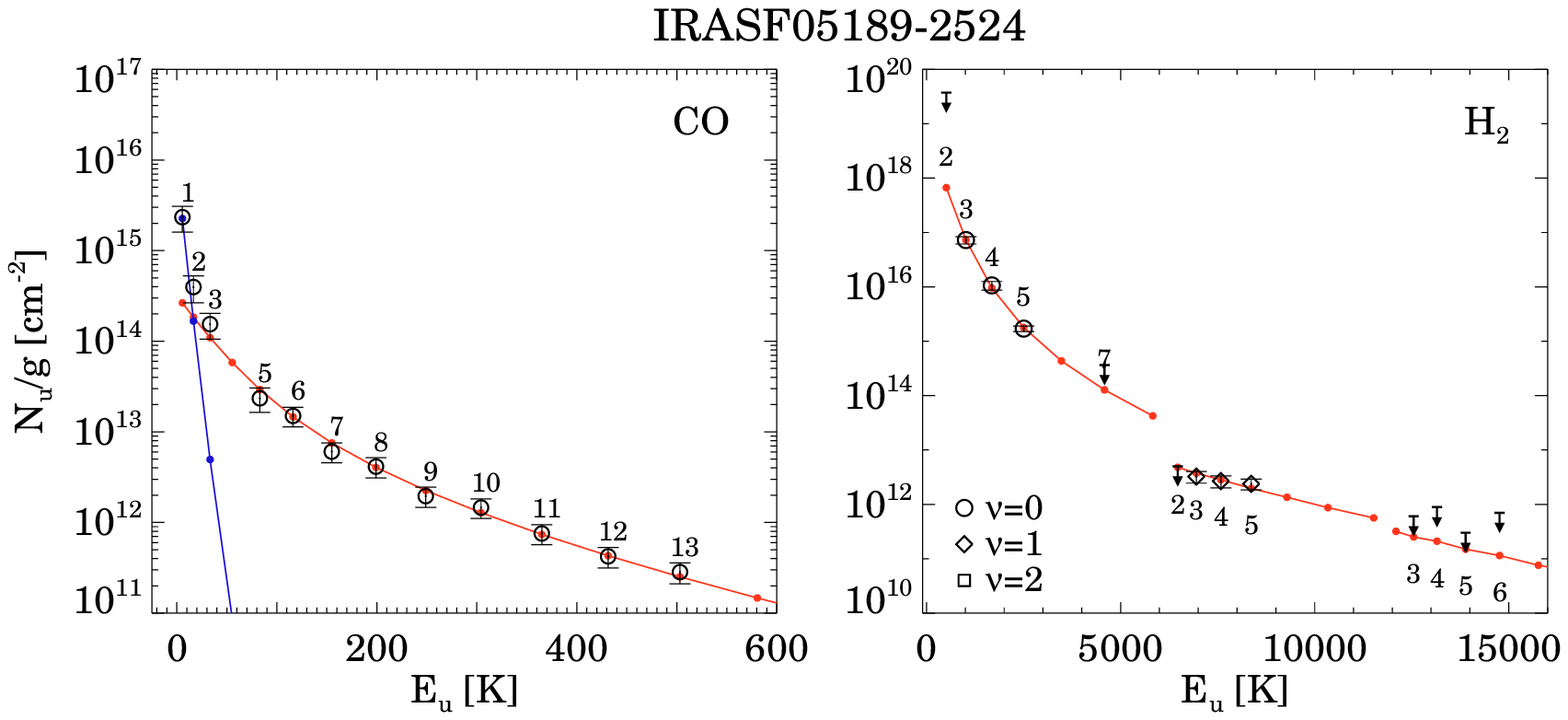}
\includegraphics[width=0.49\textheight]{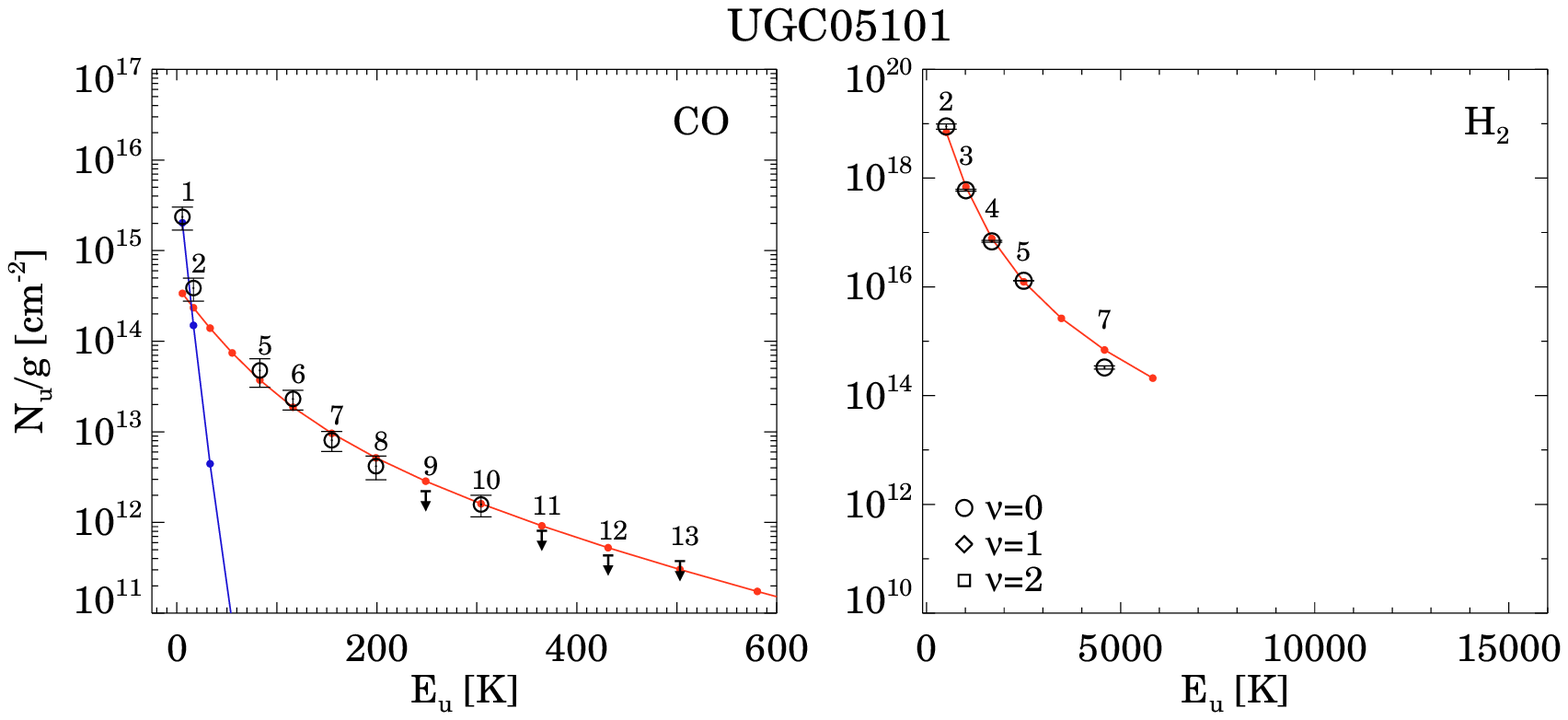}
\includegraphics[width=0.49\textheight]{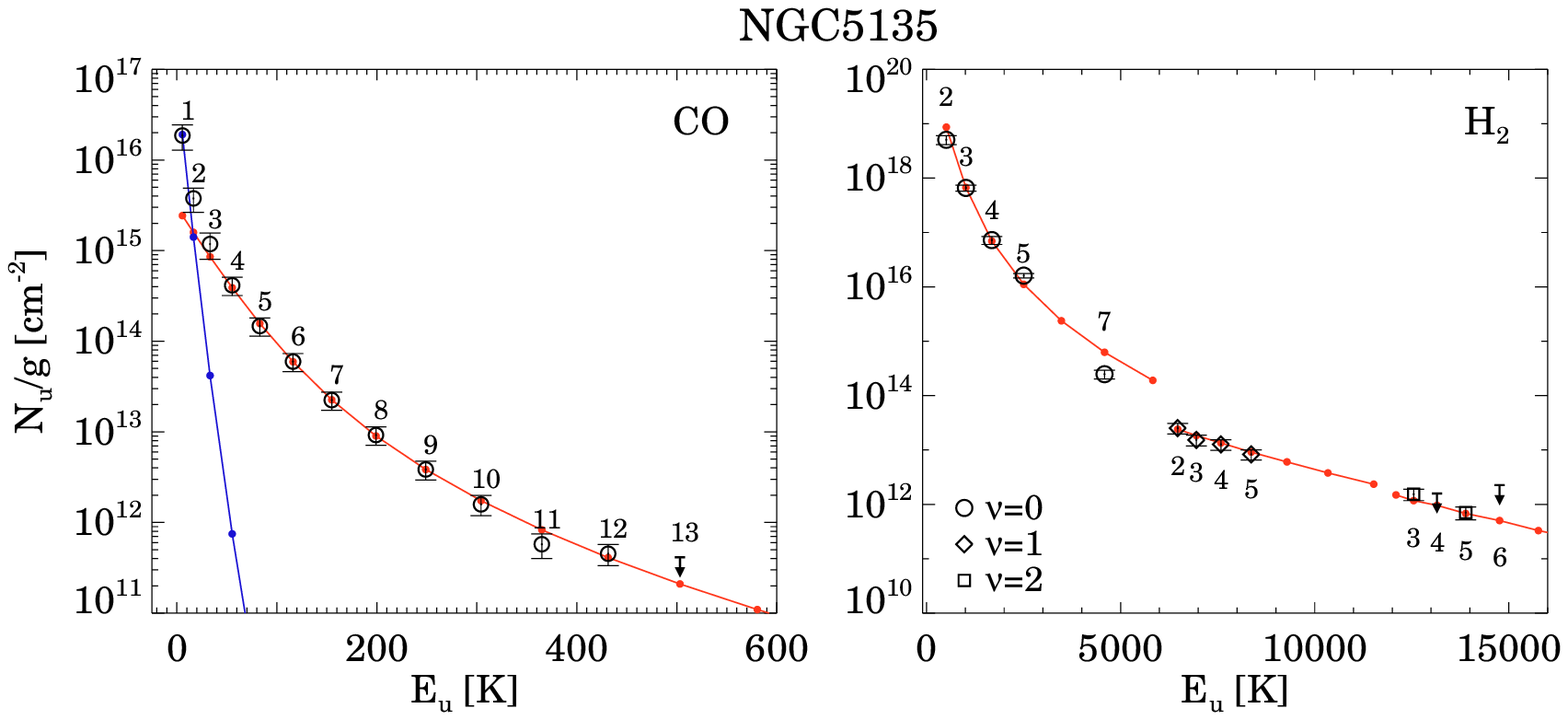}
\includegraphics[width=0.49\textheight]{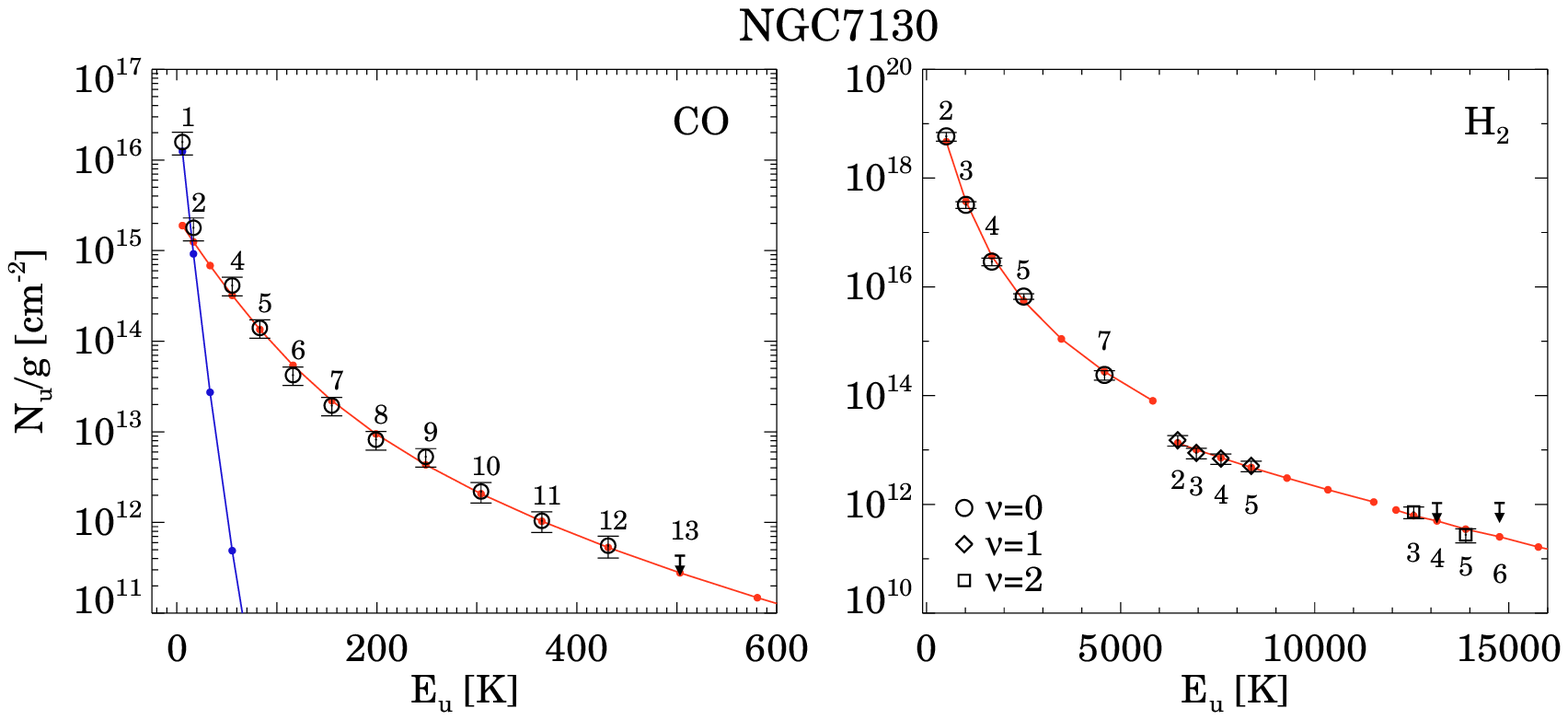}
\includegraphics[width=0.49\textheight]{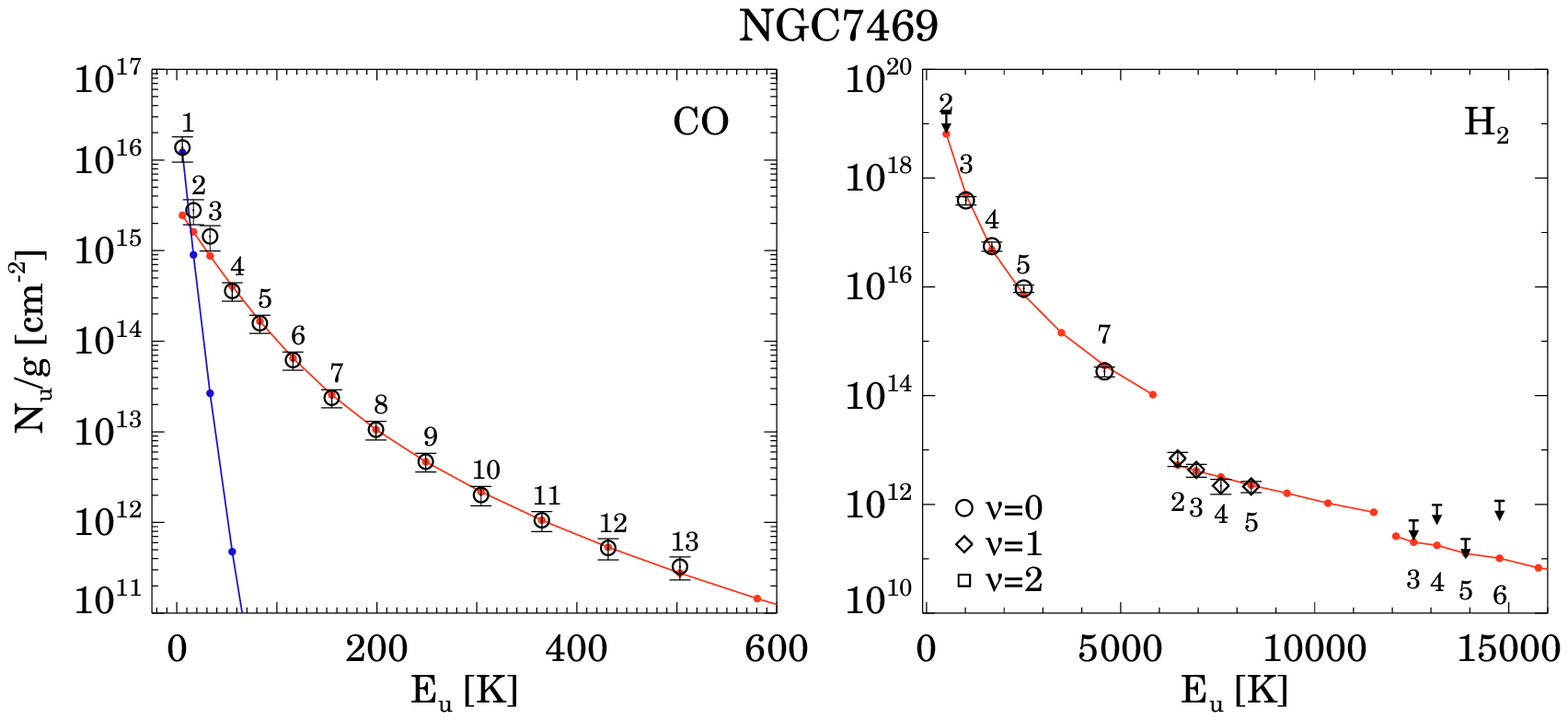}
\caption{CO and \Hmol\ rotational diagrams. The red (blue) line is the best-fit model for the warm (cold) component. The number close to the data points is the $J$ quantum number of the level. For \Hmol, circles correspond to the vibrational level $\nu=0$, diamonds to $\nu=1$, and squares to $\nu=2$,
\label{fig:powerlaw_fit}}
\end{minipage}}
\end{figure*}

We used this to estimate a lower limit for $n_{\rm H_2}$ given that the CO abundance should be lower or equal to that of C. Assuming the solar atomic C abundance (2.4$\times$10$^{-4}$; \citealt{Lodders2003}), the lower limit for $n_{\rm H_2}$ is $\sim$10$^{4.0 \pm 0.5}$\,cm$^{-3}$, depending on the galaxy. In the case of sub-solar abundances in these galaxies, the $n_{\rm H_2}$ upper limit would be slightly higher.
Similarly, we can obtain a lower limit for the CO abundance by setting $n_{\rm H_2}$ to a value above the CO critical density (e.g., 10$^{6}$\,cm$^{-3}$). The minimum CO abundance ranges between 10$^{-5}$ and 10$^{-6}$. In Table \ref{tab:mol_gas}, we list the best-fit parameters for $n_{\rm H_2}=10^{4.5}$\,cm$^{-3}$ and $n_{\rm H_2}=10^{6}$\,cm$^{-3}$. 
We plot the best-fit models in Figure \ref{fig:powerlaw_fit}.

The $n_{\rm H}$ value depends mostly on the ratio between the intensities of the rotational and near-IR ro-vibrational \Hmol\ lines. For our galaxies we obtain $n_{\rm H}$ values between 10$^{3}$--10$^{4}$\,cm$^{-3}$. The critical densities for collisions with H of the near-IR \Hmol\ transitions are $n_{\rm H}\sim 10^{4}-10^{5}$\,cm$^{-3}$ at 1500\,K, much lower than the critical densities for collisions with \Hmol. Therefore, a relatively small abundance of atomic H can enhance the luminosity of the near-IR transitions. For the assumed $n_{\rm H_2}=10^{4.5}$ and 10$^6$\,cm$^{-3}$ the ro-vibrational near-IR emission would be one or two orders of magnitude lower than the observed if we did not include collisions with H in our model.

\section{Cold-to-warm molecular gas ratio}\label{s:cold-to-warm_ratio}

To estimate the cold molecular gas mass, we used the CO to \Hmol\ conversion factor from \citet{Leroy2011}, $N_{\rm c}({\rm H_2})$(cm$^{-2}$) = $2\times 10^{20} I_{{\rm CO} J =1-0}$\,(K\,\kms). We compare this cold molecular gas column density with the warm molecular gas column density obtained from the fit of the CO and \Hmol\ SLEDs (Table \ref{tab:mol_gas}). The warm molecular gas fraction ranges between 1 and 20\% for the models with $n_{\rm H_2}=10^{4.5}$\,cm$^{-3}$ and between 15 and $>100$\% for the models with $n_{\rm H_2}=10^{6}$\,cm$^{-3}$. This suggests that for the two galaxies where the warm column density exceeds the cold column density (UGC~05101 and NGC~5135) the model with $n_{\rm H_2}=10^{6}$\,cm$^{-3}$ overestimates the $N_{\rm w}({\rm H_2})$. Thus, in the following we use the $n_{\rm H_2}=10^{4.5}$\,cm$^{-3}$ model for these two galaxies. Moreover, these are the only two sources where the $\chi^2$ is lower for the $n_{\rm H_2}=10^{4.5}$\,cm$^{-3}$ model.

For our sample of IR bright galaxies it might be adequate to use the lower CO to \Hmol\ conversion factor $N_{\rm H_2}$(cm$^{-2}$) = $0.5\times 10^{20} I_{{\rm CO} J =1-0}$\,(K\,\kms) measured by \citet{Downes1998} in active galaxies. Therefore, the cold-to-warm ratio could be lower by a factor of 4. 

\section{Molecular gas heating}\label{s:heating}

In the previous sections, we showed that the CO and \Hmol\ SLEDs are compatible with a broken power-law temperature distribution, however, we did not discuss how the molecular gas is heated. Three main mechanisms are invoked to explain the heating of the molecular gas: UV radiation, shocks, and X-ray and cosmic rays. In this section, we compare the molecular gas temperature distribution expected in photodissociation region (PDRs; UV radiation) and shocks regions with the observed temperature distribution.

X-ray and cosmic ray heating can be important for AGNs, however, in this work we analyze active galaxies with strong star-formation (see Table \ref{tab:sample}).
The AGN contribution to the mid-$J$ CO emission seems to be small compared to that of SF (e.g., \citealt{Pereira2013}). Likewise, SF dominates the \Hmol\ emission in LIRGs (e.g., \citealt{Roussel07, Pereira2010}). For these reasons, we consider that X-ray heating should not contribute significantly in the integrated spectra of these galaxies.

\begin{figure*}
\centering
\includegraphics[width=1.8\columnwidth]{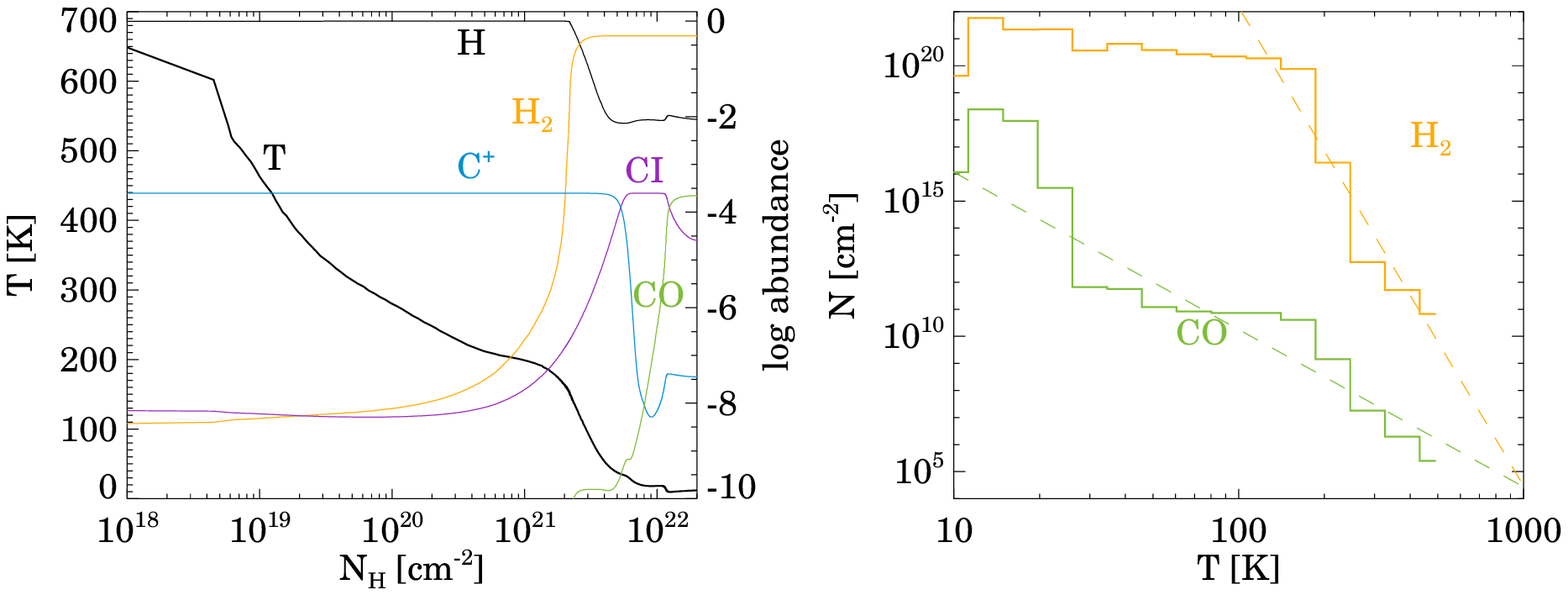}
\caption{Left panel: Chemical structure and temperature of a PDR with $G_0=10^4$ and $n_{\rm H}=10^4$\,cm$^{-3}$. The thick black line is the gas temperature (left axis). The other lines represent the abundances (left axis) of H (black), \Hmol\ (orange), C$^+$ (blue), \ion{C}{I} (violet), and CO (green). Right panel: \Hmol\ (orange) and CO (green) column densities as a function of the temperature. The dashed lines are the best power-law fits to the distribution (see Section \ref{ss:cloudy_pdr}). \label{fig:pdr_cloudy}}
\end{figure*}

\subsection{UV radiation: PDRs}\label{ss:cloudy_pdr}

We used the radiative transfer code \textsc{Cloudy} version 13.02 \citep{Ferland2013} to estimate the gas temperature distribution of PDRs. We used a constant density slab model illuminated by the interstellar radiation field of \citet{Black1987b}. We created a grid of models with $n_{\rm H}$ between $10^4$ and $10^6$\,cm$^{-3}$ and radiation field intensity between $10^1$ and $10^4$\,$G_{\rm 0}$, where $G_{\rm 0}=1.6\times 10^{-3}$\,erg\,cm$^{-2}$\,s$^{-1}$ is the intensity integrated between 6 and 13.6\,eV. Since we are interested only in the PDR region we extinguished the input spectra to remove the ionizing radiation.

We extracted the CO and \Hmol\ abundances and kinetic temperatures as a function of the column density from the output of the models, which we used to compute the temperature distribution of CO and \Hmol. Then we fitted these distributions with a power-law function. In the fit we only included gas with $T>20$\,K and $T>150$\,K, for CO and \Hmol, respectively, because lower temperature gas does not produce significant emission of the observed transitions.

In Figure \ref{fig:pdr_cloudy}, we show an example \textsc{Cloudy} PDR model and the fit to the temperature distribution. We find that, in general, \Hmol\ distributions are steeper than those of CO. The power-law index\footnote{This index refers to the $\beta$ of the relation d$N\slash$d$T\sim T^{-\beta}$.} for CO is $\sim$3--10 for $n_{\rm H_2}=10^{4.5}$\,cm$^{-3}$, and $\sim$5--7 for $n_{\rm H_2}=10^{6}$\,cm$^{-3}$, and for \Hmol\ $\beta$ ranges between $\sim$10 and 15. That is, these $\beta$ values are higher than the values derived from our fits to the observed SLEDs (Section \ref{s:sled_fit}).

\subsection{Shocks}\label{ss:shocks}

To calculate the molecular gas temperature distribution in shocks we used a simple model. We assumed that molecular gas is heated at a constant rate up to a certain temperature, which depends on the shock velocity and is high enough to produce \Hmol\ emission, and after that it is cooled radiatively by the dominant cooling molecule.

The main coolant of warm molecular gas is \Hmol, CO, or H$_2$O depending on the density and temperature of the gas \citep{Neufeld1993}. For our galaxies, the H$_2$O transitions are weaker than the CO transitions for those with similar $E_{\rm up}$ values. That is, H$_2$O cooling does not seem to dominate the cooling for the average conditions in these galaxies, and therefore, we do not consider H$_2$O in our model.

First, we calculated the non-LTE cooling functions of CO and \Hmol\ as a function of the gas density and kinetic temperature using the grid of models described in Section \ref{ss:single_model}. For the \Hmol\ cooling function we do not include the effect of collisional excitation by H since it would only affect the cooling of high-temperature gas through ro-vibrational emission of \Hmol.

The temperature dependence of these cooling functions can be approximated with a power-law (e.g., \citealt{Draine1993}). Thus we fitted the calculated cooling functions using a power-law in the temperature range between 30 and 150\,K for CO, and 150 and 1000\,K for \Hmol. These are the temperature ranges where CO and \Hmol\ dominate the gas cooling \citep{Neufeld1993}.

\begin{figure}
\centering
\includegraphics[width=0.9\columnwidth]{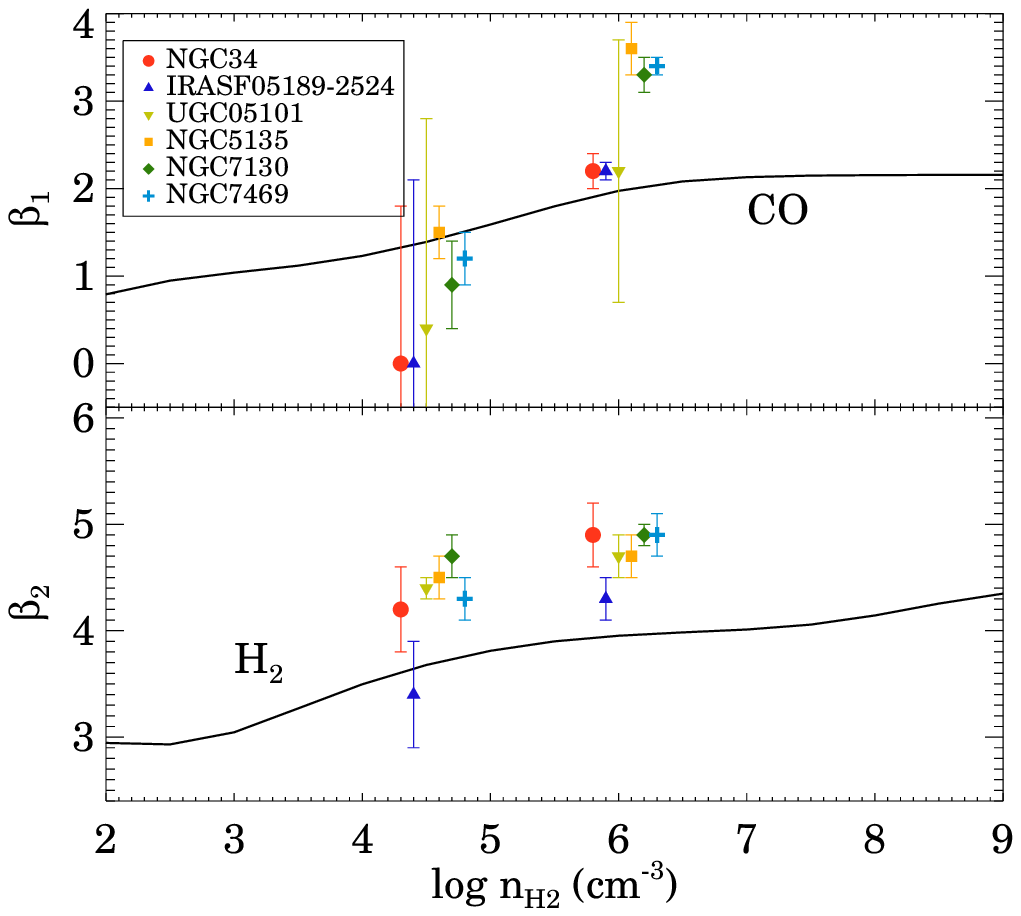}
\caption{Power-law indices of the temperature dependence of the CO (top) and \Hmol\ (bottom) non-LTE cooling functions (solid line; see Section \ref{ss:shocks}). The fits are valid for a temperature range between 30 and 150\,K for CO and 150 and 1000\,K for \Hmol. The fitted $\beta$ values form the models with $n_{\rm H_2}=10^{4.5}$\,cm$^{-3}$ and $n_{\rm H_2}=10^6$\,cm$^{-3}$ are also plotted for each galaxy. For visualization the abscissa of the data points are slightly offset. \label{fig:shocks_cooling}}
\end{figure}

Then, assuming that the molecular gas cools radiatively after being heated according to these fitted cooling functions (i.e, d$U\slash$ d$t \sim -T^\gamma$, where $U=5\slash 2\,kT$ is the internal kinetic energy), the observed temperature distribution of the molecular gas is d$N\slash$d$T \sim T^{-\gamma}$. That is, the $\beta$ we fitted for the SLEDs should be compared with $\gamma$. The power-law indices, $\gamma$, for CO and \Hmol\ as a function of the density are shown in Figure \ref{fig:shocks_cooling}.

\subsection{Comparison with the observations}

In our fits, we obtained that the power-law index varies between 1.2 and 3.5 for CO and between 4.3 and 4.9 for \Hmol. In general, these values are lower than those expected in PDR, but higher than those expected from shocks (see Table \ref{tab:mol_gas} and Figure \ref{fig:shocks_cooling}). Therefore, the combination of PDRs and shocks are needed to explain the intermediate power-law indices.

The CO emission is compatible with that predicted by shock models only for NGC~5135\footnote{We do not consider UGC~05101 because its $\beta_1$ uncertainty, 2.4, is very large.}, whose best-fit model has $n_{\rm H_2}=10^{4.5}$\,cm$^{-3}$. This result, however, strongly depends on the adopted density. For instance, for the model with $n_{\rm H_2}=10^{6}$\,cm$^{-3}$, which provides only a slightly worse fit to the CO SLED of NGC~5135, the CO emission would need a PDR component.

If we combine these PDR and shock models, the PDR SLEDs would dominate the lower-$J$ CO transitions because of their steeper power-law index, while the contribution from shocks would be dominant for higher-$J$ CO transitions. Moreover, only gas with $T>100$\,K produces \Hmol\ emission so, depending on the $G_0$ values of the PDRs and the shock velocities, the contributions of PDRs and shocks may differ in the CO and \Hmol\ SLEDs.

Our sample includes three spiral galaxies (NGC~5135, NGC~7130, and NGC~7469) and three major mergers (NGC~34, IRASF~05189--2524, and UGC~05101). The latter have lower $\beta_1$ values for their CO SLEDs (Figure \ref{fig:shocks_cooling}) suggesting that the relative contribution from shocks to the CO emission in these mergers is larger than in the spirals. For the \Hmol\ emission only IRASF~05189--2524 has a $\beta_2$ value lower than the rest of galaxies, so in this source, shocks able to heat the molecular gas up to $T>100$\,K could be important. Also, this galaxy is the most luminous in our sample and the galaxy with the higher AGN contribution (Table \ref{tab:sample}), so the bright AGN might affect its CO and \Hmol\ emissions.

\subsection{Comparison with two-component models}

The CO SLEDs of two of our galaxies (UGC~05101 and NGC~7130) were presented by \citet{Pereira2013}. In that paper, a model consisting of two components was used to fit the data: a cold component that contributed mainly to the \CO1 emission, and a warm component with variable \Hmol\ density and kinetic temperature.

Once corrected for the different beam sizes used, the beam-averaged CO column densities derived by both methods agree within the uncertainties, although, this could be a particular case due to the specific SLEDs of these two galaxies.

The interpretation of the \nht\ and \tkin\ of the two-component model (TCM) is not so straightforward.
The high kinetic temperature needed by the TCM reveals the presence of warm molecular gas, but the exact temperature value is not necessarily a representative temperature of the molecular gas.
The obtained \nht\ and \tkin\ are biased toward the conditions in the most luminous regions. Actually, the luminosity of each CO transition, for a given molecular mass, depends on \nht\ and \tkin. So, the derived \nht\ and \tkin\ will depend on the unknown distribution of densities and temperatures of the molecular gas. This problem is partially solved assuming a functional form for the d$N\slash$d$T$, as we adopted in our models.

In addition, \nht\ and \tkin\ are correlated (lower \nht\ and higher \tkin, and higher \nht\ and lower \tkin\ produce similar CO SLEDs) and this adds an extra uncertainty to the derived values. This is equivalent to the uncertainty due to the $\beta$-\nht\ correlation for our models (see Section \ref{s:sled_fit}).

Although the TCM provides useful information, when the analyzed molecular transitions comprise a wide range of excitation temperatures, the need for additional components for the SLED fitting can make the modeling contrived. Instead, the power-law models used here could better represent the integrated temperature distribution of the molecular gas in a galaxy.

\section{Conclusions}\label{s:conclusions}

We have studied the integrated CO and \Hmol\ emission of six local IR bright galaxies using non-LTE models. Assuming a broken power-law distribution for the molecular gas temperatures, our model reproduces both the CO SLED (from $J_{\rm up}=1$ to $J_{\rm up}=13$) and the \Hmol\ SLED ($J_{\rm up}\leq 7$ for the lowest three vibrational levels) in our sample of galaxies.

The main findings of this work are summarized in the following:

\begin{enumerate}
\item With a single power-law temperature distribution it is not possible to fit simultaneously the CO and \Hmol\ SLEDs. The \Hmol\ SLEDs have a much steeper power-law index ($\beta_2\sim 4-5$) than the CO SLED ($\beta_1\sim 1-3$). This is the expected behavior for the temperature distributions in PDRs and shocks.

\item We found that for most of the galaxies, the models with $n_{\rm H_2}=10^6$\,cm$^{-3}$ provide the best fit to the observed data, thus the majority of the transitions are close to LTE. The minimum acceptable density for the warm gas is $\sim n_{\rm H_2}=10^{4 \pm 0.5}$\,cm$^{-3}$, lower densities would imply CO abundances higher than the atomic C abundance. Likewise, we obtained that the minimum CO abundance in the warm gas is $x_{\rm CO}\sim 10^{-6}-10^{-5}$ assuming LTE conditions.

\item The column densities of the warm molecular gas ($T>20$\,K) represents between 10 and 100\,\% of the molecular gas traced by the \CO1 transition.

\item We used PDR and shock models to determine the excitation mechanism of the molecular gas. Our models show that the temperature distributions are steeper for PDRs than for shocks. We also found that the temperature distribution of the warmest gas ($T>100$\,K) emitting in \Hmol\ is steeper than that of coldest gas ($T>30$\,K), which produces the mid-$J$ CO emission for both PDR and shocks models. This is because of the different main coolant of the warm and cold molecular gas (\Hmol, and CO, respectively).

\item Neither PDR nor shocks alone can explain the derived temperature distributions. A combination of both is needed to reproduce the observed SLEDs. In this case, the lower-$J$ CO transitions would be dominated by PDR emission, whereas the higher-$J$ CO transitions would be dominated by shocks.

\item The three major mergers among our targets (NGC~34, IRASF~05189--2524, and UGC~05101) have shallower temperature distributions for CO than the other three spirals.
This suggests that the relative contribution of shocks to the heating of warm molecular gas ($T<100$\,K) in these major mergers is higher than in the other three spirals.

\item For only one of the mergers, IRASF~05189--2524, the shallower \Hmol\ temperature distribution (hot molecular gas) suggests that the relative importance of shocks is high. This galaxy also has a bright AGN that dominates the bolometric luminosity, which can contribute to the molecular gas heating.
For the other two mergers, the \Hmol\ temperature distribution is similar to that of the spiral galaxies. Therefore the shocks producing the extra contribution to the CO emission in these mergers are not able to heat the molecular gas to temperatures higher than 100\,K, which would be necessary to see the \Hmol\ emission.

\end{enumerate}

\begin{acknowledgements}

We thank anonymous referee for comments that improved the paper.
This work was funded by the Agenzia Spaziale Italiana (ASI) under contract I/005/11/0. JPL acknowledge support from the Spanish Plan Nacional de Astronom\'ia y Astrof\'isica AYA2010-21161-C02-01 and AYA2012-32295.
SPIRE has been developed by a consortium of institutes led by Cardiff Univ. (UK) and including: Univ. Lethbridge (Canada); NAOC (China); CEA, LAM (France); IFSI, Univ. Padua (Italy); IAC (Spain); Stockholm Observatory (Sweden); Imperial College London, RAL, UCL-MSSL, UKATC, Univ. Sussex (UK); and Caltech, JPL, NHSC, Univ. Colorado (USA). This development has been supported by national funding agencies: CSA (Canada); NAOC (China); CEA, CNES, CNRS (France); ASI (Italy); MCINN (Spain); SNSB (Sweden); STFC, UKSA (UK); and NASA (USA).
This research has made use of the NASA/IPAC Extragalactic Database (NED) which is operated by the Jet Propulsion Laboratory, California Institute of Technology, under contract with the National Aeronautics and Space Administration.

\end{acknowledgements}

\end{document}